\documentclass[a4paper,prl,twocolumn,showpacs,superscriptaddress,groupedaddress]{revtex4}
\usepackage{ae}
\usepackage{physics}
\usepackage[T1]{fontenc}
\usepackage[ansinew]{inputenc}
\usepackage{amsmath}
\usepackage{amssymb}
\usepackage[caption=false]{subfig}
\usepackage{multirow}
\usepackage{array}
\usepackage[]{graphicx}
\usepackage{wrapfig}
\usepackage{makecell}
\usepackage{xparse}
\usepackage{dcolumn}
\usepackage{color}
\usepackage[colorlinks]{hyperref}
\usepackage{lscape}
\graphicspath{ {./images1/} }

\begin{document}
	\title{\huge{More assistance of entanglement, less rounds of classical communication}}
	\author{\large\bfseries Atanu Bhunia}
	\email{atanu.bhunia31@gmail.com}
	\affiliation{Department of Applied Mathematics, University of Calcutta, 92, A.P.C. Road, Kolkata- 700009, India}
	\author{\large\bfseries Indranil Biswas}
	\email{indranilbiswas74@gmail.com}
	\affiliation{Department of Applied Mathematics, University of Calcutta, 92, A.P.C. Road, Kolkata- 700009, India}
	\author{\large\bfseries Indrani Chattopadhyay}
	\email{icappmath@caluniv.ac.in}
	\affiliation{Department of Applied Mathematics, University of Calcutta, 92, A.P.C. Road, Kolkata- 700009, India}
	\author{\large\bfseries Debasis Sarkar}
	\email{dsarkar1x@gmail.com, dsappmath@caluniv.ac.in}
	\affiliation{Department of Applied Mathematics, University of Calcutta, 92, A.P.C. Road, Kolkata- 700009, India}
	
	\begin{abstract}
		{\normalsize\textbf{Abstract}}\\
		Classical communication plays a crucial role to distinguish locally a class of quantum states. Despite considerable advances, we have very little knowledge about the number of measurement and communication rounds needed to implement a discrimination task by local quantum operations and classical communications (in short, LOCC). In this letter, we are able to show the relation between round numbers with the local discrimination of a set of pure bipartite orthogonal quantum states. To demonstrate the possible strong dependence on the round numbers, we consider a class of orthogonal product states in $d\otimes d$, which require at least $2d-2$ round of classical communications. Curiously the round number can be reduced to $d$ by the assistance of one-ebit of entanglement as resource and can be reduced further by assistance of more entanglement. We are also able to show that the number of LOCC rounds needed for a discrimination task may depend on the amount of entanglement assistances.
	\end{abstract}
	\date{\today}
	\maketitle
{\bfseries 1.\;Introduction}\\
Nonlocal properties of quantum systems have a class exclusive from Bell nonlocality. Specifically, when a set of orthogonal quantum states cannot be perfectly distinguished by local operations and classical communications (LOCC), it reflects another nonlocal feature of quantum physics  \cite{Bennett1999}. Local distinguishability of quantum states refers to the task of distinguishing a state from a set of prespecified orthogonal states shared among parties separated by arbitrary distances and LOCC being the only legitimate class of operations \cite{Bennett1999,bennett1996,popescu2001,xin2008,Walgate2000,Virmani,Ghosh2001,Groisman,Walgate2002,Divincinzo,Horodecki2003,Fan2004,Ghosh2004,Nathanson2005,Watrous2005,Niset2006,Ye2007,Fan2007,Runyo2007,somsubhro2009,Feng2009,Runyo2010,Yu2012,Yang2013,Zhang2014,somsubhro2009(1),somsubhro2010,yu2014,somsubhro2014,somsubhro2016}. The nonlocality of sets of orthogonal quantum states can be used for various practical purposes, such as, data hiding, quantum secret sharing and so on. The study of local distinguishability of orthogonal quantum states and exploring their relationship between quantum nonlocality and entanglement received considerable attention in the past two decades \cite{Zhang2015,Wang2015,Chen2015,Yang2015,Zhang2016,Xu2016(2),Zhang2016(1),Xu2016(1),Halder2019strong nonlocality,Halder2019peres set,Xzhang2017,Xu2017,Wang2017,Cohen2008,Zhang2016(3),somsubhro2018,Halder2018,bhunia2020,bhunia2022}.\\

In quantum information processing, one of the most important physical scenario occurs when a multipartite system is distributed among different parties separated by arbitrary distances. The parties perform multiple rounds of local measurements on their respective subsystems, and each time globally broadcasting their measurement outcomes. Other parties are required to choose their measurement setups depending on the outcomes and continue the process as required. This class of operations is known as LOCC. From an experimental perspective, LOCC operations have a natural attraction since local quantum measurements are much easier to perform on a system than their nonlocal counterpart. And on an even more fundamental level, LOCC is linked to the very notion of entanglement, as entanglement is precisely the multipartite correlations that cannot be generated by LOCC. However, despite this general feature, the class of LOCC is still not satisfactorily understood. One largely overlooked the question of how the number of measurement and communication rounds allowed in an LOCC process that affects what tasks the parties are able to perform. In other words, what is the cost of the LOCC round number to accomplish a given task? Here we are asking for the number of times the parties must make a local measurement and use the classical channel to communicate their results. If the channel has some finite capacity, then this question generalizes the question of minimum classical communication cost in performing some LOCC tasks, a vitally important issue in its own right. Thus, the LOCC round number can be seen as a cost for both classical communications and quantum operations.\\

There are relatively few studies conducted on the round number. Bennett et. al., have proven that two-way LOCC is strictly more powerful than just one-way LOCC\cite{bennett1996}. On the other hand, for entanglement manipulation of pure bipartite states, Lo and Popescu \cite{popescu2001} showed that two-way communications are equivalent to one-way communications and one-way communications are provably better than no communication. For the task of distinguishing states, Xin and Duan have constructed a collection of states in $m\otimes n$ systems that needed at least $2\;\min\{m,n\}-2$ rounds of classical communications in order to be perfectly distinguished\cite{xin2008}. These findings demonstrate that the exact relationship between the round number and task achievability is a highly nontrivial issue, and in fact, contains some surprising results.\\

In this letter, we study the effect of classical communications for a local discrimination task. We show that in general many rounds of classical communications are necessary. We demonstrate this result by constructing a class of $d\otimes d$ pure orthogonal states, which requires at least $2d-2$ rounds of classical communications to achieve a perfect local discrimination. In some sense, our result exhibit that two way classical communications can effectively increase the local distinguishability. Furthermore, we show that the round number of the discrimination task can be brought down by the assistance of entanglement. Interestingly we observe that the round number can be reduced to $ d $ by the support of one-ebit of entanglement as resource and it can be decreased further by using more resources. Throughout this letter, we do not normalize states and operators for simplicity. Every bipartite pure state can be written as $|\psi\rangle=\sum_{i, j} m_{i, j}|i\rangle|j\rangle \in \mathbb{C}^{m} \otimes \mathbb{C}^{n}$, where $|i\rangle$ and $|j\rangle$ are the computational bases of $\mathbb{C}^{m}$ and $\mathbb{C}^{n}$, respectively. There exists a one to one correspondence between the state $|\psi\rangle$ and the $m \times n$ matrix $M=\left(m_{i j}\right)$. If $\operatorname{rank}(M)=1$, then $|\psi\rangle$ is a product state, and if $\operatorname{rank}(M)>1$ then $|\psi\rangle$ is an entangled state. Also $\left\langle\psi_{1} \mid \psi_{2}\right\rangle=\operatorname{Tr}\left(M_{1}^{\dagger} M_{2}\right)$, where $\left\langle\psi_{1} \mid \psi_{2}\right\rangle$ is the inner product of $\left|\psi_{1}\right\rangle$ and $\left|\psi_{2}\right\rangle$. Now, firstly we will review some definitions which we will use in the following discussions.

$Definition 1.$\cite{Halder2018} If all the POVM elements of a measurement structure corresponding to a discrimination task of a given set of states are proportional to the identity matrix, then such a measurement is not useful to extract information for this task and is called a $trivial\;measurement$. On the other hand, if not all POVM elements of a measurement are proportional to the identity matrix then the measurement	is said to be a $nontrivial\; measurement$.\\

$Definition 2.$\cite{Halder2018} Consider a measurement to distinguish a fixed set of pairwise orthogonal quantum states. After performing that measurement, if the postmeasurement states are also pairwise orthogonal to each other then such a measurement is said to be an $orthogonality-preserving\;\;measurement$(OPM).\\

$Definition 3.$ The number of classical communications round required for a discrimination task means the number of times the parties globally broadcast their measurement outcomes after performing the local measurement on their respective subsystems.\\\\
{\bfseries 2.\;Distinguishability by minimum classical round}\\
Here we construct a set of orthogonal pure product states which require a minimum rounds of classical communications for the respective discrimination task. For better understanding, we first provide an example in ${\mathbb{C}}^{6}\bigotimes{\mathbb{C}}^{6}$ and generalize the result to higher dimensions. We represent here a quantum state $\ket{i+\overline{i+1}}\ket{j}$ by a rectangle, where $\ket{i\pm\overline{i+1}}=\frac{1}{\sqrt{2}}(\ket{i}\pm\ket{i+1}),$ for integer $i$. \\

\emph{\textbf{Proposition 1}}. The $36$ states in $6\otimes6$,
	\begin{multline}
		$$
		\;\;\;\;\;\;\;\;\;\;\;|a\pm b\rangle =\frac{1}{\sqrt{2}}(|a\rangle \pm|b\rangle), 0 \leq a<b \leq 6,\\
		\left|\phi_{1,2}\right\rangle =|0\rangle_{A}|0\pm1\rangle_{B}, \;\;\;\;\;\;\;
		\left|\phi_{3,4}\right\rangle =|0\rangle_{A}|2\pm3\rangle_{B}, \\
		\left|\phi_{5,6}\right\rangle =|0\rangle_{A}|4\pm 5\rangle_{B}, \;\;\;\;\;\;\;
		\left|\phi_{7,8}\right\rangle =|1\pm2\rangle_{A}|0\rangle_{B}, \\
		\left|\phi_{9,10}\right\rangle =|3\pm4\rangle_{A}|0\rangle_{B}, \;\;\;\;\;\;\;\;\;\;\;\;\;
		\left|\phi_{11}\right\rangle =|5\rangle_{A}|0\rangle_{B}, \\
		\;\;\;\left|\phi_{12,13}\right\rangle \left.=\left.|1\rangle_{A}\right|1\pm2\right\rangle_{B}, \;\;\;\;
		\left|\phi_{14,15}\right\rangle =|1\rangle_{A}|3\pm 4\rangle_{B}, \\
		\;\;\;\left|\phi_{16}\right\rangle =|1\rangle_{A}|5\rangle_{B},\;\;\;\;\;\;\;\;\;\;\;\;\;\;\;\;
		\left|\phi_{17,18}\right\rangle =|2\pm3\rangle_{A}|1\rangle_{B}, \\
		\;\;\left|\phi_{19,20}\right\rangle =|4\pm5\rangle_{A}|1\rangle_{B},\;\;\;\;\;\;\;
		\left|\phi_{21,22}\right\rangle =|2\rangle_{A}|2\pm3\rangle_{B}, \\
		\;\;\left|\phi_{23,24}\right\rangle =|2\rangle_{A}|4\pm5\rangle_{B} ,\;\;\;\;\;\;\;
		\left|\phi_{25,26}\right\rangle =|3\pm4\rangle_{A}|2\rangle_{B},\\
		\left|\phi_{27}\right\rangle =|5\rangle_{A}|2\rangle_{B}, \;\;\;\;\;\;\;\;\;\;
		\left|\phi_{28,29}\right\rangle =|3\rangle_{A}|3\pm4\rangle_{B} ,\\
		\left|\phi_{30}\right\rangle =|3\rangle_{A}|5\rangle_{B},\;\;\;\;\;\;\;\;\;\;
		\left|\phi_{31,32}\right\rangle =|4\pm5\rangle_{A}|3\rangle_{B}, \\
		\left|\phi_{33,34}\right\rangle =|4\rangle_{A}|4\pm5\rangle_{B} ,\;\;\;\;\;\;\;\;\;\;
		\left|\phi_{35}\right\rangle =|5\rangle_{A}|4\rangle_{B},\\
		\left|\phi_{36}\right\rangle =|5\rangle_{A}|5\rangle_{B},\\
		$$
		\label{1}
	\end{multline}
	need at least ten rounds of classical communications to be distinguishable by LOCC.\\

	\begin{figure}[h!]
		\centering
		\includegraphics[width=0.45\textwidth]{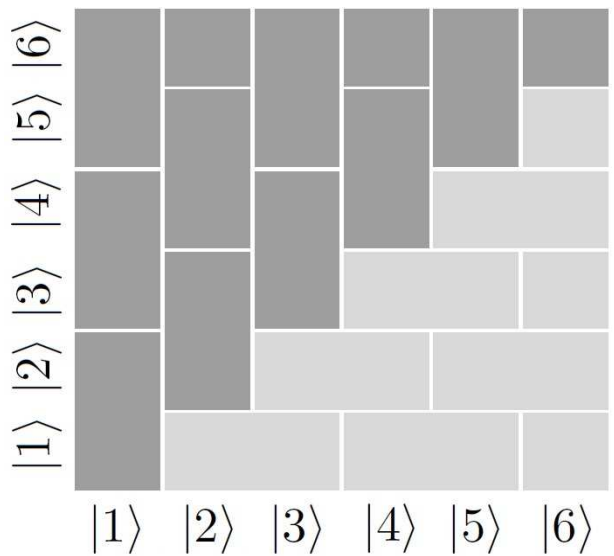}
		\caption{Product states representation in ${\mathbb{C}}^{6}\bigotimes{\mathbb{C}}^{6}$. Here, we represent a quantum state $\ket{i+\overline{i+1}
			}\ket{j}$ by a rectangle where $\ket{i\pm\overline{i+1}
			}=\frac{1}{\sqrt{2}}(\ket{i}\pm\ket{i+1}),$ for integer $i$.
		}
	\end{figure}

	\emph{\textbf{Proof:}} Suppose Alice goes first, and let $A_{m}$ denote Alice's POVM operator with outcome
	$m$ such that the postmeasurement $\quad$ states $\quad\left\{A_{m} \otimes I_{B} \left|\phi_{i}\right\rangle, i=\right.$
	$1, \ldots, 36\}$ should be mutually orthogonal. Because $a_{i j}=0$ is necessary and sufficient for $a_{j i}=0, i<j$, we will only show $a_{i j}=0, i<j$. Now, considering the states $\left|\phi_{1,12}\right\rangle$, we have $\left\langle 0\left|A_{m}\right| 1\right\rangle_{A} \left\langle 0+1|1+2\right\rangle_{B}=0;$ which implies, $a_{01}=a_{10}=0 .$ In the same way, for the states $\left|\phi_{3,21}\right\rangle,\left|\phi_{3,28}\right\rangle,\left|\phi_{5,33}\right\rangle$ and $\left|\phi_{5,36}\right\rangle$,
	we have $a_{02}=a_{20}=0, a_{03}=a_{30}=0, a_{04}=a_{40}=0$ and $a_{05}=a_{50}=0$, respectively. Similarly, if we choose the states $\left|\phi_{12,21}\right\rangle,\left|\phi_{14,28}\right\rangle,\left|\phi_{14,33}\right\rangle$ and $\left|\phi_{16,36}\right\rangle$, we obtain $a_{12}=a_{21}=0, a_{13}=a_{31}=0, a_{14}=a_{41}=0$ and $a_{15}=a_{51}=0$, respectively. Now considering the states $\left|\phi_{21,28}\right\rangle$, we have $\left\langle 2\left|A_{m}\right| 3\right\rangle_{A} \left\langle 2+3|3+4\right\rangle_{B}=0.$ Which imply $a_{23}=a_{32}=0 .$ In a similar manner by considering $\left|\phi_{23,33}\right\rangle,\left|\phi_{23,36}\right\rangle,\left|\phi_{28,33}\right\rangle$,$\left|\phi_{30,36}\right\rangle$ and $\left|\phi_{33,36}\right\rangle$, we have $a_{24}=a_{42}=0, a_{25}=a_{52}=0, a_{34}=a_{43}=0, a_{35}=a_{53}=0$ and $a_{45}=a_{54}=0$, respectively. Therefore, $A_{m}$ is diagonal and $A_{m}=\operatorname{diag}\left(\alpha_{0}, \alpha_{1}, \ldots, \alpha_{5}\right)$.\\
	Next considering $\left|\phi_{7,8}\right\rangle, \quad$ we $\quad$ get $\quad\left\langle1+2\left|A_{m}\right|1-2\right\rangle_{A}\langle 0|0\rangle_{B}=0,$ i.e.,
	$\left\langle1\left|A_{m}\right|1\right\rangle-\left\langle 2\left|E_{m}\right|2\right\rangle=0$. Thus, $a_{11}=a_{22}$. For the states
	$\left|\phi_{9,10}\right\rangle,\left|\phi_{17,18}\right\rangle,$ and $\left|\phi_{19,20}\right\rangle,$ we finally get $a_{11}=a_{22}=\cdots=a_{55}.$ Therefore, $A_{m}$
	$=\operatorname{diag}\left(\alpha_{0}, \beta, \beta \ldots, \beta\right).$
	If possible, let us assume that $\alpha_{0}\neq0$ and $\beta\neq0$. Then after Alice's measurement, Bob should do a nontrivial operation on his own subsystem according to Alice's result. We denote $B_{n}$ as Bob's operator. As we have discussed above, by choosing suitable pair of states we can conclude that all the off-diagonal elements of $B_{n}$ are equal to $0$. Similarly, for the diagonal elements as we have discussed above, if we consider the states $\left|\phi_{1,2}\right\rangle,\left|\phi_{3,4}\right\rangle,\left|\phi_{5,6}\right\rangle,\left|\phi_{12,13}\right\rangle$ and $\left|\phi_{14,15}\right\rangle,$ we finally get, $b_{00}=b_{11}=\cdots=b_{55} .$ Therefore, $B_{n}$ is proportional to the identity operator, i.e., $B_{n}=\gamma_{0}I$, which is the trivial operator and this contradicts our assumption. So, either $\alpha_{0}=0$ or $\beta=0.$ Notice that this result also suggests us that these states cannot be distinguished locally if Bob goes first. Now it is clear that if Alice goes first with a diagonal operator, i.e., $\alpha_{0}=\beta=1$, then the above set of states cannot be distinguished. So, Alice has to do non-trivial measurement first and this only happens when any one of $\alpha_{0}$, $\beta$ is not equal to zero. For that Alice only has two outcome measurement operators: $\quad A_{1}=\operatorname{diag}(1,0, \ldots, 0)$ and $A_{2}$ $=\operatorname{diag}(0,1,1 \ldots, 1)$. If the outcome $A_1$ click, Bob is able to distinguish the remaining states by projecting onto $\left|0\pm1\right\rangle,\left|2\pm3\right\rangle$ and $\left|4\pm5\right\rangle$. If the measurement outcome is $A_2$, it will isolate the remaining 30 states. Therefore the system is now $5\otimes6 .$
	It is then Bob's turn to do measurement. Following the method we used above, we can prove that Bob's measurement must be $E_{1}=\operatorname{diag}(1,0, \ldots, 0)$ and $E_{2}$ $=\operatorname{diag}(0,1, \ldots, 1)$. By induction, we find the number of rounds needed for distinguishing is $10.$ This completes the proof.\(\blacksquare \)\\
	
	Obviously, the states of the above set constitute a basis. It is not possible to distinguish the above class of states with lesser number of rounds. Also it is noted that if we omit or add some states into the set, it will change the minimum bound of round number. Next we generalize the result for arbitrary large dimensions.\\
	
	\emph{\textbf{Proposition 2.}} The $d^{2}$ states in $d\otimes d$ system, where $d$ is even,
	\begin{multline}
		$$
		\;\;\;\;\;\;\;\;\;\;\;\;\;|a\pm b\rangle =\frac{1}{\sqrt{2}}(|a\rangle \pm|b\rangle), 0 \leq a<b, \\
		\left|\phi_{i+1, i+2}\right\rangle =|0\rangle_{A}|i\pm(i+1)\rangle_{B}, i=0,2, \ldots, d-2,\;\;\;\;\;\;\;\;\;\\
		\left|\phi_{d+i, d+i+1}\right\rangle =|i\pm(i+1)\rangle_{A}|0\rangle_{B}, i=1,3, \ldots, d-3,\;\;\;\;\;\\
		\left|\phi_{2d-1}\right\rangle =|d-1\rangle_{A}|0\rangle_{B},\;\;\;\;\;\;\;\;\;\;\;\;\;\;\;\;\;\;\;\;\;\;\;\;\;\;\;\;\;\;\;\;\;\;\;\;\;\;\;\;\;\;\;\;\;\;\;\;\;\;\\
		\left|\phi_{2d+i-1,2d+i}\right\rangle =|1\rangle_{A}|i\pm(i+1)\rangle_{B}, i=1,3, \ldots, d-3,\;\;\;\\
		\left|\phi_{3d-2}\right\rangle =|1\rangle_{A}|d-1\rangle_{B},\;\;\;\;\;\;\;\;\;\;\;\;\;\;\;\;\;\;\;\;\;\;\;\;\;\;\;\;\;\;\;\;\;\;\;\;\;\;\;\;\;\;\;\;\;\;\;\;\;\;\\
		\left|\phi_{3d+i-3,3d+i-2}\right\rangle =|i\pm (i+1)\rangle_{A}|1\rangle_{B}, i=2,4, \ldots, d-2,\\
		\left|\phi_{4d+i-5,4d+i-4}\right\rangle =|2\rangle_{A}|i\pm (i+1)\rangle_{B}, i=2,4, \ldots, d-2,\\
		\left|\phi_{5d+i-8,5d+i-7}\right\rangle =|i\pm (i+1)\rangle_{A}|2\rangle_{B}, i=3,5, \ldots, d-3,\\
		\left|\phi_{6d-9}\right\rangle =|d-1\rangle_{A}|2\rangle_{B},\;\;\;\;\;\;\;\;\;\;\;\;\;\;\;\;\;\;\;\;\;\;\;\;\;\;\;\;\;\;\;\;\;\;\;\;\;\;\;\;\;\;\;\;\;\;\;\;\;\;\;\\
		\left|\phi_{6d+i-11,6d+i-10}\right\rangle =|3\rangle_{A}|i \pm(i+1)\rangle_{B}, i=3,5, \ldots, d-3,\\
		\left|\phi_{7d-12}\right\rangle =|3\rangle_{A}|d-1\rangle_{B},\;\;\;\;\;\;\;\;\;\;\;\;\;\;\;\;\;\;\;\;\;\;\;\;\;\;\;\;\;\;\;\;\;\;\;\;\;\;\;\;\;\;\;\;\;\;\;\;\;\;\\
		\left|\phi_{7d+i-15,7d+i-14}\right\rangle =|i\pm (i+1)\rangle_{A}|3\rangle_{B}, i=4,6, \ldots, d-2,\\
		\left|\phi_{8d+i-19,8d+i-18}\right\rangle =|4\rangle_{A}|i\pm (i+1)\rangle_{B}, i=4,6, \ldots, d-2,\\
		\left|\phi_{9d+i-24,9d+i-23}\right\rangle =|i\pm (i+1)\rangle_{A}|4\rangle_{B}, i=5,7, \ldots, d-3,\\
		\left|\phi_{10d-25}\right\rangle =|d-1\rangle_{A}|4\rangle_{B},\;\;\;\;\;\;\;\;\;\;\;\;\;\;\;\;\;\;\;\;\;\;\;\;\;\;\;\;\;\;\;\;\;\;\;\;\;\;\;\;\;\;\;\;\;\;\;\;\;\;\\
		\left|\phi_{10d+i-29,10d+i-28}\right\rangle =|5\rangle_{A}|i\pm (i+1)\rangle_{B}, i=5,7, \ldots, d-3,\\
		\left|\phi_{11d-30}\right\rangle =|5\rangle_{A}|d-1\rangle_{B},\;\;\;\;\;\;\;\;\;\;\;\;\;\;\;\;\;\;\;\;\;\;\;\;\;\;\;\;\;\;\;\;\;\;\;\;\;\;\;\;\;\;\;\;\;\;\;\;\;\;\\
		\vdots\;\;\;\;\;\;\;\;\;\;\;\;\;\;\;\;\;\;\;\;\;\;\;\;\;\;\;\;\;\;\;\;\;\;\;\;\;\;\;\;\;\;\;\;\;\;\;\;\;\;\\
		\left|\phi_{d^2-5,d^2-4}\right\rangle =|(d-2)\pm(d-1)\rangle_{A}|d-3\rangle_{B},\;\;\;\;\;\;\;\;\;\;\;\;\;\;\;\;\;\;\;\;\;\;\;\;\;\;\;\;\;\;\;\;\;\;\;\;\\
		\left|\phi_{d^2-3,d^2-2}\right\rangle =|d-2\rangle_{A}|(d-2)\pm(d-1)\rangle_{B},\;\;\;\;\;\;\;\;\;\;\;\;\;\;\;\;\;\;\;\;\;\;\;\;\;\;\;\;\;\;\;\;\;\;\;\;\\
		\left|\phi_{d^2-1}\right\rangle =|d-1\rangle_{A}|d-2\rangle_{B},\;\;\;\;\;\;\;\;\;\;\;\;\;\;\;\;\;\;\;\;\;\;\;\;\;\;\;\;\;\;\;\;\;\;\;\;\;\;\;\;\;\;\;\;\;\;\;\;\;\;\\
		\left|\phi_{d^2}\right\rangle =|d-1\rangle_{A}|d-1\rangle_{B},\;\;\;\;\;\;\;\;\;\;\;\;\;\;\;\;\;\;\;\;\;\;\;\;\;\;\;\;\;\;\;\;\;\;\;\;\;\;\;\;\;\;\;\;\;\;\;\;\;\;\\
		$$
		\label{2}
	\end{multline}
	need at least $2d-2$ rounds classical communications to be distinguishable by LOCC.\\
	
	\emph{\textbf{Proof:}} See supplementary information\cite{Supplementary information} for explicit description of the proof.

	By the above construction it is not very difficult to find a set which requires a fixed amount of round number of classical communications for its discrimination task. The main factor which plays an important role in this structure is the quantum superposition. In the next discussion, we construct an entanglement assisted discrimination protocol for the above set of states with lesser number of communication rounds.\\\\
	{\bfseries 3.\;Reducing classical round by one-ebit}\\ We now consider the discrimination protocol of the above class of states by using entanglement as a resource. \\
	 
    \emph{\textbf{Proposition 3.}} The set of states (\ref{1}) needs only six rounds of communications for its local discrimination task by consuming one copy of $2\otimes 2$ maximally entangled state as a resource.\\
	
	\emph{\textbf{Proof:}} First of all we assume that one-ebit of entanglement shared between Alice, Bob be $|\psi\rangle_{ab}$. Therefore the initial state shared among them is $\left|\phi\right\rangle_{AB}\otimes\left|\psi\right\rangle_{ab}$,
	where $\left|\phi\right\rangle$ is one of the state from (\ref{1}).\\

	\begin{figure}[h!]
		\centering
		\includegraphics[width=0.45\textwidth]{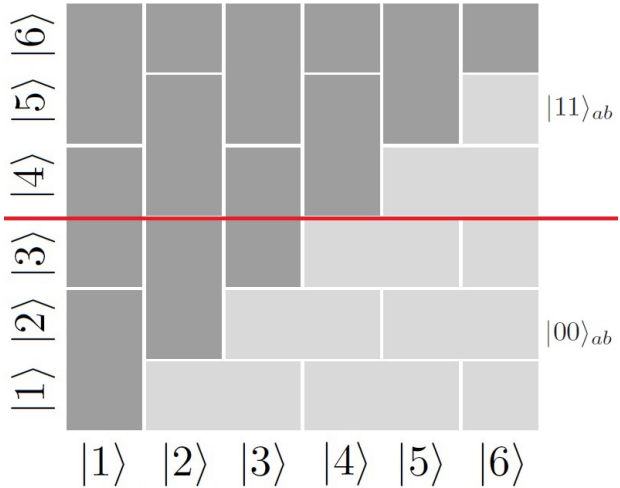}
		\caption{Product states representation in ${\mathbb{C}}^{6}\bigotimes{\mathbb{C}}^{6}$ following Bob's first measurement with outcome B1. The labels on the right column correspond to Alice's and Bob's assisted systems.
		}
	\end{figure}

	\emph{Round 1.} Bob performs a measurement
	\begin{multline*}
		$$
		\mathcal{B}\equiv\{B_1:=\mathbb{P}\left[(|0\rangle,|1\rangle,|2\rangle)_{B} ;|0\rangle_{b}\right]+ \\ \mathbb{P}\left[(|3\rangle,|4\rangle,|5\rangle)_{B} ;|1\rangle_{b}\right],\\
		B_2:=\mathbb{I}-B_1\},
		$$ 
	\end{multline*}
	where $\mathbb{P}(\cdot)$ represents the projection operator.
    Later on Alice and Bob do some sequence of measurements to distinguish locally the class of states (\ref{1}). The complete description of proof is in the supplementary information\cite{Supplementary information}.

	Next we generalize the result for arbitrary large dimensions.\\
	\emph{\textbf{Proposition 4.}} The set of states (\ref{2}) needs only $d$ rounds of communications for its local discrimination task by consuming one copy of $2\otimes 2$ maximally entangled state as a resource.\\
	\emph{\textbf{Proof:}} See supplementary information\cite{Supplementary information} for complete description of the proof.\\\\
	{\bfseries 4.\;Reducing classical round by 2-ebits}\\
	\label{A4}
	Now we will present a method to locally distinguish
	the above class of orthogonal product states in $d\otimes d$ with multiple
	copies of $2\otimes 2$ maximally entangled states. We consider multicopy resource assisted discrimination of the nonlocal set. Recall that this set of operations strictly includes the set of LOCC operations. Our result, however, establishes that, given two copies of the Bell state the number of classical communication rounds of local discrimination task can be reduced further.
	\\
	
	\emph{\textbf{Proposition5.}} The set of states (\ref{1}) needs only four rounds of communications for its local discrimination task by consuming two copies of $2\otimes 2$ maximally entangled states as a resource.\\
	
	\emph{\textbf{Proof:}} First of all let us assume that The state with 2-ebits of entanglement shared between Alice, Bob be $|\psi_1\rangle_{a_1b_1} \otimes|\psi_2\rangle_{a_2b_2}$ where each of $|\psi_1\rangle_{a_1b_1}$ and $|\psi_2\rangle_{a_2b_2}$ are of one-ebit entanglement. Therefore the initial state shared among them is $$\left|\phi\right\rangle_{AB}\otimes\left|\psi_1\right\rangle_{a_1b_1}\otimes\left|\psi_2\right\rangle_{a_2b_2}$$
	where $\left|\phi\right\rangle$ is one of the state from (\ref{1}).\\
	\emph{Round 1.} Bob performs a measurement
	\begin{multline*}
		$$
		\mathcal{B}\equiv\left\{B_1:=\mathbb{P}\left[(|0\rangle,|1\rangle,|2\rangle)_{B} ;|0\rangle_{b}\right]+\mathbb{P}\left[(|3\rangle,|4\rangle,|5\rangle)_{B} ;|1\rangle_{b}\right]\right.,\\\;\;\;\;\;\;\;\;\;\;\;\;\;\;\;\;\;\;\;\;\;\;\;\;\;\;\;\;\;\;\;\;\;\;\;\;\;\;\;\;\;\;\;\;\;\;\;\;\;\;\;\;\;\;\;\;\;\;\;\;\;\;\;\;
		B_2:=\mathbb{I}-B_1\}
		$$\\
		$$
		\mathcal{C}\equiv\left\{C_1:=\mathbb{P}\left[(|0\rangle,|4\rangle,|5\rangle)_{B} ;|0\rangle_{b}\right]+\mathbb{P}\left[(|1\rangle,|2\rangle,|3\rangle)_{B} ;|1\rangle_{b}\right]\right.,\;\;\;\;\;\;\;\\\;\;\;\;\;\;\;\;\;\;\;\;\;\;
		C_2:=\mathbb{I}-C_1\}
		$$
	\end{multline*}
Later on Alice and Bob do some sequences of measurements to distinguish the class of states (\ref{1}). The complete description of proof is in the supplementary information\cite{Supplementary information}.\\

We have presented a different distinguishing method which uses two or more low-dimensional entanglement resources instead of a high-dimensional entanglement resource. We think that our method is more efficient and saves resources.\\
\emph{\textbf{Proposition 6.}} The set of states (\ref{2}) needs only $d-2$ rounds of communications for its discrimination task by consuming two copies of $2\otimes 2$ maximally entangled states as a resource.\\
\emph{\textbf{Proof:}} See supplementary information\cite{Supplementary information} for complete description of the proof.

The round of classical communications can be further decreased by using more amount of entanglement resource for this discrimination task. In this particular task it can be checked that by using 3-ebits of entanglement resource the round number can be bring down from $d$ to $d-4$.\\\\
{\bfseries 5.\;Conclusions}\\ In this letter, we have investigated the number of measurement and communication rounds needed to implement a discrimination task by local quantum operations and classical communications (LOCC). In particular, we have constructed a special set of $d\otimes d$, states which require at least $2d-2$ rounds of classical communications for perfect discrimination. Our result indicates that classical communication plays a crucial role in local discrimination. Next  with entanglement as a resource to distinguish orthogonal quantum states, we present a method based on multiple copies of low-dimensional entanglement resources instead of a high-dimensional entanglement resource. Remarkably we have observed that  the amount of classical communications can be reduced further with the help of entanglement assistance. The results can lead to a better understanding of the relationship between classical communications and entanglement resources. However, there are still some questions worth looking for. Firstly, is it possible to extend the whole scenario to multipartite case and what will be the entanglement resource that gives advantage. Secondly, by using lesser amount of entanglement resource is it possible to get the same advantages in discrimination task.\\\\
{\bfseries Acknowledgements}\\
The authors  AB and IB acknowledge the support from UGC, India. The authors IC and DS acknowledge the work as part of QUest initiatives by DST India.
		
\section{\large Supplementary information}
{\large\textbf{Proof of Proposition 2.}}\\
Suppose Alice goes first, and let $A_{m}$ denote Alice's POVM operator with outcome
m. For discrimination, the postmeasurement $\quad$ states $\quad\left\{A_{m} \otimes I_{B} \left|\phi_{i}\right\rangle, i=\right.$
$1, \ldots, d^2\}$ should be mutually orthogonal. Because $a_{i j}=0$ is necessary and sufficient for $a_{j i}=0, i<j$, we will only show $a_{i j}=0, i<j$, in the following. Considering the states $\left|\phi_{1,2d}\right\rangle$, we have, $\left\langle 0\left|A_{m}\right| 1\right\rangle_{A} \left\langle 0+1|1+2\right\rangle_{B}=0.$ Thus, $a_{01}=a_{10}=0 .$ In the same way, for the states $\left|\phi_{3,4d-3}\right\rangle,\left|\phi_{3,6d-8}\right\rangle,\ldots,\left|\phi_{d-1,d^2}\right\rangle$, we have, $a_{02}=a_{20}=0, a_{03}=a_{30}=0$,\ldots , $a_{0(d-1)}=a_{(d-1)0}=0$, respectively. Similarly, if we choose the states $\left|\phi_{2d,4d-3}\right\rangle,\left|\phi_{2d+2,6d-8}\right\rangle,\ldots,\left|\phi_{3d-2,d^2}\right\rangle$,
we have $a_{12}=a_{21}=0, a_{13}=a_{31}=0,\ldots,a_{1(d-1)}=a_{(d-1)1}=0$, respectively. In a similar manner by considering suitable choice of states we can show that all the off-diagonal elements of $A_{m}$ becomes zero. Therefore, $A_{m}$ is diagonal and $A_{m}=\operatorname{diag}\left(\alpha_{0}, \alpha_{1}, \ldots, \alpha_{d-1}\right)$.\\
Now considering $\left|\phi_{d+1,d+2}\right\rangle, \quad$ we $\quad$ get $\quad\left\langle1+2\left|A_{m}\right|1-2\right\rangle_{A}\langle 0|0\rangle_{B}=0, \quad$ i.e.,
$\left\langle1\left|A_{m}\right|1\right\rangle-\left\langle 2\left|E_{m}\right|2\right\rangle=0$. Thus, $a_{11}=a_{22}$. By using the states
$\left|\phi_{d+3,d+4}\right\rangle,\left|\phi_{d+5,d+6}\right\rangle,\ldots\left|\phi_{2d-3,2d-2}\right\rangle,\left|\phi_{3d-1,3d}\right\rangle,\\ \left|\phi_{3d+1,3d+2}\right\rangle,\ldots\left|\phi_{4d-5,4d-4}\right\rangle,$ we finally get $a_{11}=a_{22}=\cdots=a_{(d-1)(d-1)} .$ Therefore, $A_{m}$
$=\operatorname{diag}\left(\alpha_{0}, \beta, \beta \ldots, \beta\right) .$
If possible let us assume that $\alpha_{0}\neq0$ and $\beta\neq0$. Then after Alice's measurement, Bob should do a nontrivial operation on his own system according to Alice's result. We denote $B_{n}$ as Bob's operator. As we have discussed above, by choosing suitable pair of states we can conclude that all the off-diagonal element of $B_{n}$ is equal to 0. Similarly for the diagonal element as we have discussed above, if we take $\left|\phi_{1,2}\right\rangle,\left|\phi_{3,4}\right\rangle,\ldots\left|\phi_{d-1,d}\right\rangle,\left|\phi_{2d,2d+1}\right\rangle,\left|\phi_{2d+2,2d+3}\right\rangle,\ldots\\\left|\phi_{3d-4,3d-3}\right\rangle,$ we finally get $b_{00}=b_{11}=\cdots=b_{(d-1)(d-1)} .$ Therefore $B_{n}$ is proportional to the identity operator, i.e., $B_{n}=\gamma_{0}I$. Which is trivial operator and this contradicts our assumption. So, either $\alpha_{0}=0$ or $\beta=0 .$
Notice that this result also suggests that these states cannot be distinguished if Bob goes first. Now it is clear that if Alice goes first with a diagonal operator, i.e., $\alpha_{0}=\beta=1.$ Then the above set of states cannot be distinguished. So, Alice has to do non-trivial measurement first and this will only happen when any one of $\alpha_{0}$, $\beta$ not equal to zero. For that Alice only has two outcome measurement operators: $\quad A_{1}=\operatorname{diag}(1,0, \ldots, 0)$ and $A_{2}$ $=\operatorname{diag}(0,1,1 \ldots, 1)$. If the outcome $A_1$ click, Bob will able to distinguish the left states by projecting onto $\left|0\pm1\right\rangle,\left|2\pm3\right\rangle$$\ldots$ $\left|(d-2)\pm(d-1)\right\rangle$. If the measurement outcome is $A_2$, it isolates the remaining $d^2$ states. Therefore the system is now $(d-1)\otimes d.$
It is then Bob's turn to do measurement. Following the method we used above, we can similarly prove that Bob's measurement must be $E_{1}=\operatorname{diag}(1,0, \ldots, 0)$ and $E_{2}$ $=\operatorname{diag}(0,1, \ldots, 1)$. By induction, we find the number of rounds needed for distinguishing is $2d-2 .$ This completes the proof.\\\\
{\large\textbf{Proof of Proposition 3.}}\\
First of all let us assume that one-ebit of entanglement shared between Alice, Bob be $|\psi\rangle_{ab}$. Therefore the initial states shared among them is $\left|\phi\right\rangle_{AB}\otimes\left|\psi\right\rangle_{ab}$,
Where $\left|\phi\right\rangle$ is one of the state from set of equations (1).\\
{\bfseries Round 1.} Bob performs a measurement,
\begin{multline*}
$$
\mathcal{B}\equiv\{B_1:=\mathbb{P}\left[(|0\rangle,|1\rangle,|2\rangle)_{B} ;|0\rangle_{b}\right]+ \\ \mathbb{P}\left[(|3\rangle,|4\rangle,|5\rangle)_{B} ;|1\rangle_{b}\right],\\
B_2:=\mathbb{I}-B_1\}
$$
\end{multline*}
where $\mathbb{P}(\cdot)$ represents the projection operator (we will use this notation many times in the following discussions).
Suppose the outcome corresponding to $B_1$ clicks. The resulting postmeasurement states are therefore
\begin{multline*}
$$
\left|\phi_{1,2}\right\rangle \rightarrow\left|0\rangle_{A}|0 \pm 1\rangle_{B}|00\rangle_{ab}\right.=\left|\phi\prime_{1,2}\right\rangle,\;\;\;\;\;\;\;\;\;\;\;\;\;\;\;\;\;\;\;\;\;\;\;\;\;\;\;\;\;\;\;\;\;\;\;\;\;\\
\left|\phi_{3,4}\right\rangle \rightarrow\left|0\rangle_{A}|2\rangle_{B}|00\rangle_{ab}\right.\pm\left|0\rangle_{A}|3\rangle_{B}|11\rangle_{ab}\right.=\left|\phi\prime_{3,4}\right\rangle,\;\;\;\;\;\;\;\;\;\;\;\;\;\;\;\;\;\;\;\;\;\;\;\;\;\;\;\;\;\;\;\;\;\;\;\;\;\\
\left|\phi_{5,6}\right\rangle \rightarrow\left|0\rangle_{A}|4\pm5\rangle_{B}|11\rangle_{ab}\right.=\left|\phi\prime_{5,6}\right\rangle,\;\;\;\;\;\;\;\;\;\;\;\;\;\;\;\;\;\;\;\;\;\;\;\;\;\;\;\;\;\;\;\\
\left|\phi_{7,8}\right\rangle \rightarrow\left|1\pm2\rangle_{A}|0\rangle_{B}|00\rangle_{ab}\right.=\left|\phi\prime_{7,8}\right\rangle,\;\;\;\;\;\;\;\;\;\;\;\;\;\;\;\;\;\;\;\;\;\;\;\;\;\;\;\;\;\;\\
\left|\phi_{9,10}\right\rangle \rightarrow\left|3\pm4\rangle_{A}|0\rangle_{B}|00\rangle_{ab}\right.=\left|\phi\prime_{9,10}\right\rangle,\;\;\;\;\;\;\;\;\;\;\;\;\;\;\;\;\;\;\;\;\;\;\;\;\;\;\;\;\;\\
\left|\phi_{11}\right\rangle \rightarrow\left|5\rangle_{A}|0\rangle_{B}|00\rangle_{ab}\right.=\left|\phi\prime_{11}\right\rangle,\;\;\;\;\;\;\;\;\;\;\;\;\;\;\;\;\;\;\;\;\;\;\;\;\;\;\;\;\;\;\;\;\;\;\;\;\;\\
\left|\phi_{12,13}\right\rangle \rightarrow\left|1\rangle_{A}|1\pm2\rangle_{B}|00\rangle_{ab}\right.=\left|\phi\prime_{12,13}\right\rangle,\;\;\;\;\;\;\;\;\;\;\;\;\;\;\;\;\;\;\;\;\;\;\;\;\;\;\;\;\;\;\\
\left|\phi_{14,15}\right\rangle \rightarrow\left|1\rangle_{A}|3\pm4\rangle_{B}|11\rangle_{ab}\right.=\left|\phi\prime_{14,15}\right\rangle,\;\;\;\;\;\;\;\;\;\;\;\;\;\;\;\;\;\;\;\;\;\;\;\;\;\;\;\;\;\\
\left|\phi_{16}\right\rangle \rightarrow\left|1\rangle_{A}|5\rangle_{B}|11\rangle_{ab}\right.=\left|\phi\prime_{16}\right\rangle,\;\;\;\;\;\;\;\;\;\;\;\;\;\;\;\;\;\;\;\;\;\;\;\;\;\;\;\;\;\;\;\;\;\;\;\;\;\\
\left|\phi_{17,18}\right\rangle \rightarrow\left|2\pm3\rangle_{A}|1\rangle_{B}|00\rangle_{ab}\right.=\left|\phi\prime_{17,18}\right\rangle,\;\;\;\;\;\;\;\;\;\;\;\;\;\;\;\;\;\;\;\;\;\;\;\;\;\;\;\;\;\;\\
\left|\phi_{19,20}\right\rangle \rightarrow\left|4\pm5\rangle_{A}|1\rangle_{B}|00\rangle_{ab}\right.=\left|\phi\prime_{19,20}\right\rangle,\;\;\;\;\;\;\;\;\;\;\;\;\;\;\;\;\;\;\;\;\;\;\;\;\;\;\;\;\;\\
\left|\phi_{21,22}\right\rangle \rightarrow\left|2\rangle_{A}|2\rangle_{B}|00\rangle_{ab}\right.\pm\left|2\rangle_{A}|3\rangle_{B}|11\rangle_{ab}\right.=\left|\phi\prime_{21,22}\right\rangle,\;\;\;\;\;\;\\
\left|\phi_{23,24}\right\rangle \rightarrow\left|2\rangle_{A}|4\pm5\rangle_{B}|11\rangle_{ab}\right.=\left|\phi\prime_{23,24}\right\rangle,\;\;\;\;\;\;\;\;\;\;\;\;\;\;\;\;\;\;\;\;\;\;\;\;\;\;\;\;\;\\
\left|\phi_{25,26}\right\rangle \rightarrow\left|3\pm4\rangle_{A}|2\rangle_{B}|00\rangle_{ab}\right.=\left|\phi\prime_{25,26}\right\rangle,\;\;\;\;\;\;\;\;\;\;\;\;\;\;\;\;\;\;\;\;\;\;\;\;\;\;\;\;\;\\
\left|\phi_{27}\right\rangle \rightarrow\left|5\rangle_{A}|2\rangle_{B}|00\rangle_{ab}\right.=\left|\phi\prime_{27}\right\rangle,\;\;\;\;\;\;\;\;\;\;\;\;\;\;\;\;\;\;\;\;\;\;\;\;\;\;\;\;\;\;\;\;\;\;\;\;\;\\
\left|\phi_{28,29}\right\rangle \rightarrow\left|3\rangle_{A}|3\pm4\rangle_{B}|11\rangle_{ab}\right.=\left|\phi\prime_{28,29}\right\rangle,\;\;\;\;\;\;\;\;\;\;\;\;\;\;\;\;\;\;\;\;\;\;\;\;\;\;\;\;\;\\
\left|\phi_{30}\right\rangle \rightarrow\left|3\rangle_{A}|5\rangle_{B}|11\rangle_{ab}\right.=\left|\phi\prime_{30}\right\rangle,\;\;\;\;\;\;\;\;\;\;\;\;\;\;\;\;\;\;\;\;\;\;\;\;\;\;\;\;\;\;\;\;\;\;\;\;\;\\
\left|\phi_{31,32}\right\rangle \rightarrow\left|4\pm5\rangle_{A}|3\rangle_{B}|11\rangle_{ab}\right.=\left|\phi\prime_{31,32}\right\rangle,\;\;\;\;\;\;\;\;\;\;\;\;\;\;\;\;\;\;\;\;\;\;\;\;\;\;\;\;\;\\
\left|\phi_{33,34}\right\rangle \rightarrow\left|4\rangle_{A}|4\pm5\rangle_{B}|11\rangle_{ab}\right.=\left|\phi\prime_{33,34}\right\rangle,\;\;\;\;\;\;\;\;\;\;\;\;\;\;\;\;\;\;\;\;\;\;\;\;\;\;\;\;\;\\
\left|\phi_{35}\right\rangle \rightarrow\left|5\rangle_{A}|4\rangle_{B}|11\rangle_{ab}\right.=\left|\phi\prime_{35}\right\rangle,\;\;\;\;\;\;\;\;\;\;\;\;\;\;\;\;\;\;\;\;\;\;\;\;\;\;\;\;\;\;\;\;\;\;\;\;\;\\
\left|\phi_{36}\right\rangle \rightarrow\left|5\rangle_{A}|5\rangle_{B}|11\rangle_{ab}\right.=\left|\phi\prime_{36}\right\rangle,\;\;\;\;\;\;\;\;\;\;\;\;\;\;\;\;\;\;\;\;\;\;\;\;\;\;\;\;\;\;\;\;\;\;\;\;\;\\
$$
\end{multline*}
{\bfseries Round 2.} Alice performs a measurement
\begin{multline*}
$$
\mathcal{A}^2\equiv\{A_1^2:=\mathbb{P}\left[|0\rangle_{A} ;(|0\rangle,|1\rangle)_{a}]\right.,A_2^2:=\mathbb{P}\left[|1\rangle_{A} ;|1\rangle_{a}]\right.,\\\;\;\;\;\;\;\;\;\;\;A_3^2:=\mathbb{P}\left[|3\rangle_{A} ;|1\rangle_{a}]\right.,A_4^2=\mathbb{P}\left[(|4\rangle,|5\rangle)_{A} ;|1\rangle_{a}]\right.,\\A_5^2:=\mathbb{I}-\sum A_i^2\}
$$
\end{multline*}
If the outcome corresponding to $A_1^2$ clicks. The resulting post measurement states are therefore $	\left|\phi\prime_{1,2,3,4,5,6}\right\rangle $. If the outcome  $A_2^2$ clicks, it will isolate $	\left|\phi\prime_{14,15,16}\right\rangle$. Also the outcome  $A_3^2$ will isolate $	\left|\phi\prime_{28,29,30}\right\rangle $. If the outcome $A_4^2$ clicks, it remains $	\left|\phi\prime_{31,32,33,34,35,36}\right\rangle $. If the outcome $A_5^2$ clicks, it remains $	\left|\phi\prime_{7,8,9,10,11,12,13,17,18,19,20,21,22,23,24,25,26,27}\right\rangle $. \\
{\bfseries Round 3.} Bob performs the measurement,\\
depending on the outcome $A_1^2$ of the previous round, Bob performs the measurement,
\begin{multline*}
$$
\mathcal{B}^3_{1}\equiv\{B_{11}^3:=\mathbb{P}\left[|0+1\rangle_{B} ;|0\rangle_{b}]\right.,B_{12}^3:=\mathbb{P}\left[|0-1\rangle_{B} ;|0\rangle_{b}]\right.,\\\;\;\;\;\;\;\;\;\;\;\;\;\;\;\; B_{13}^3=\mathbb{P}\left[|4+5\rangle_{B} ;|1\rangle_{b}]\right.,B_{14}^3:=\mathbb{P}\left[|4-5\rangle_{B} ;|1\rangle_{b}]\right.,\\B_{15}^3:=\mathbb{I}-\sum B_i\}
$$
\end{multline*}
Now the outcomes $B_{11}^3, B_{12}^3, B_{13}^3, B_{14}^3$ will distinguish $|\phi\prime_{1,2,5,6}\rangle$ respectively. If $B_{15}^3$ clicks, it isolates $|\phi\prime_{3,4}\rangle$.\\
Depending on $A_2^2$, Bob performs the measurement,
\begin{multline*}
$$
\mathcal{B}^3_{2}\equiv\{B_{21}^3:=\mathbb{P}\left[|3+4\rangle_{B} ;|1\rangle_{b}]\right.,B_{22}^3:=\mathbb{P}\left[|3-4\rangle_{B} ;|1\rangle_{b}]\right.,\\\;\;\;\;\;\;\;\;\;\;\;\; B_{23}^3=\mathbb{P}\left[|5\rangle_{B} ;|1\rangle_{b}]\right.\}
$$
\end{multline*}
Now the outcomes $B_{21}^3, B_{22}^3, B_{23}^3$ distinguish $|\phi\prime_{14,15,16}\rangle$ respectively. Depending on $A_3^2$, Bob performs,
\begin{multline*}
$$
\mathcal{B}^3_{3}\equiv\{B_{31}^3:=\mathbb{P}\left[|3+4\rangle_{B} ;|1\rangle_{b}]\right.,B_{32}^3:=\mathbb{P}\left[|3-4\rangle_{B} ;|1\rangle_{b}]\right.,\\\;\;\;\;\;\;\;\;\;\;\;\; B_{33}^3=\mathbb{P}\left[|5\rangle_{B} ;|1\rangle_{b}]\right.\}
$$
\end{multline*}
\begin{figure}[h!]
	\centering
	\includegraphics[width=3.3in, height=4.3in]{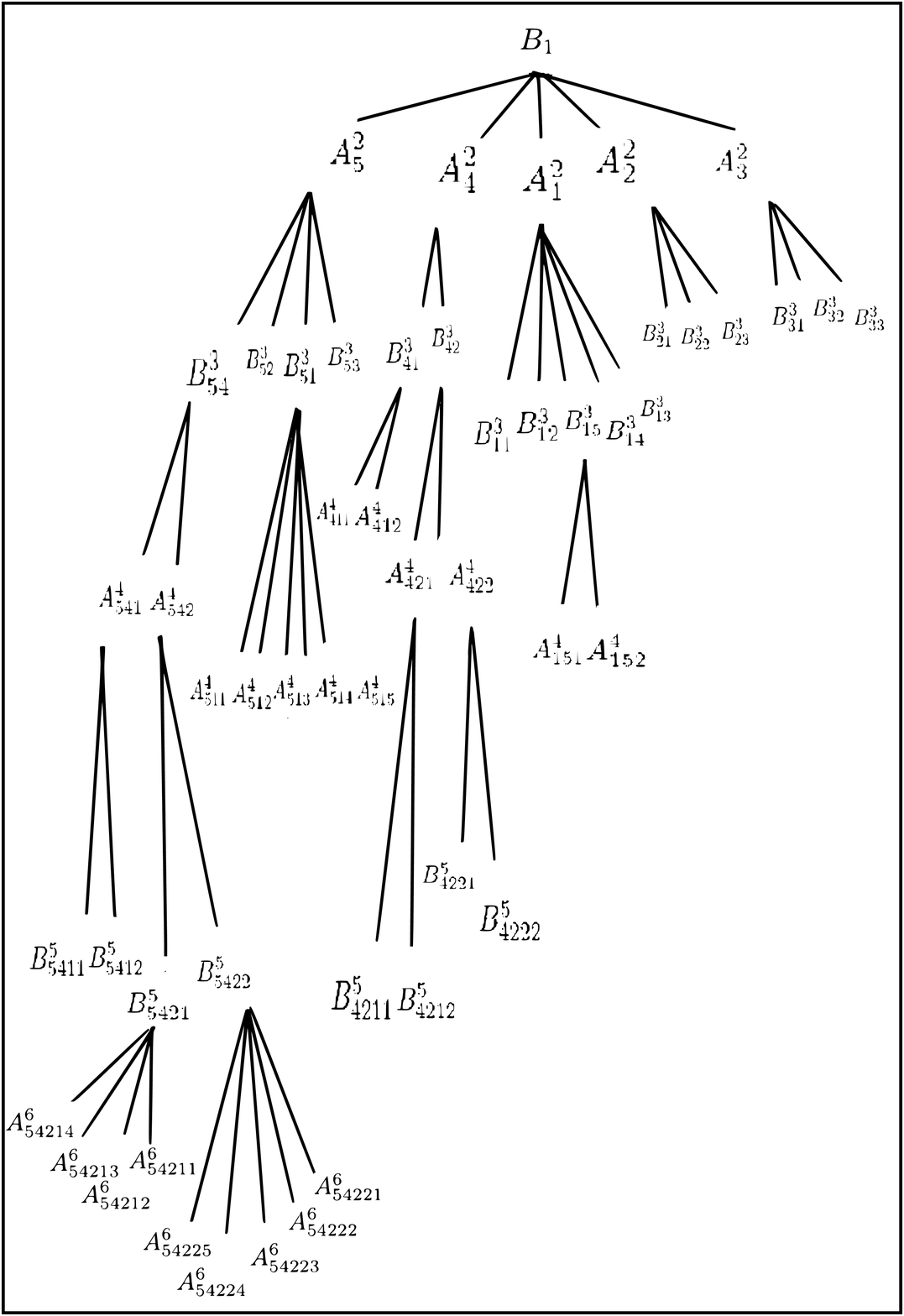}
	\caption{A six round protocol for the discrimination task of the set of states in  ${\mathbb{C}}^{6}\bigotimes{\mathbb{C}}^{6}$.
	}
\end{figure}
Now the outcomes $B_{31}^3, B_{32}^3, B_{33}^3$ distinguish $|\phi\prime_{28,29,30}\rangle$ respectively. Next depending on $A_4^2$, Bob performs,
\begin{multline*}
$$
\mathcal{B}^3_{4}\equiv\{B_{41}^3:=\mathbb{P}\left[|3\rangle_{B} ;|1\rangle_{b}]\right.,B_{42}^3:=\mathbb{I}-B_{41}^3\}\;\;\;\;\;\;\;\;\;\;\;\;
$$
\end{multline*}
If  $B_{41}^3$ clicks, it isolates $|\phi\prime_{31,32}\rangle$. Also if  $B_{42}^3$ clicks, it isolates $|\phi\prime_{33,34,35,36}\rangle$. Similarly, depending on $A_5^2$, Bob performs,
\begin{multline*}
$$
\mathcal{B}^3_{5}\equiv\{B_{51}^3=\mathbb{P}\left[|0\rangle_{B} ;|0\rangle_{b}]\right.,B_{52}^3=\mathbb{P}\left[|4+5\rangle_{B} ;|1\rangle_{b}]\right.,\\B_{53}^3=\mathbb{P}\left[|4+5\rangle_{B} ;|1\rangle_{b}]\right.,B_{54}^3:=\mathbb{I}-B_{51}^3-B_{52}^3-B_{53}^3\}
$$
\end{multline*}
The outcome $B_{51}^3$ isolates $|\phi\prime_{7,8,9,10,11}\rangle$. The outcomes $B_{52}^3, B_{53}^3$  distinguish the states $|\phi\prime_{23,24}\rangle$, respectively. Also if $B_{54}^3$ clicks, it remains $|\phi\prime_{12,13,17,18,19,20,21,22,25,26,27}\rangle$.\\
{\bfseries Round 4.}  Alice performs the measurement,\\
depending on $B_{15}^3$  Alice performs,
\begin{multline*}
$$
\mathcal{A}^4_{15}\equiv\{A_{151}^4=\mathbb{P}\left[|0\rangle_{A} ;|0+1\rangle_{a}]\right.,\\A_{152}^4=\mathbb{P}\left[|0\rangle_{A} ;|0-1\rangle_{a}]\right.\}
$$
\end{multline*}
to distinguish the states $|\phi\prime_{3,4}\rangle$. Next, depending on $B_{41}^3$  Alice performs,
\begin{multline*}
$$
\mathcal{A}^4_{41}\equiv\{A_{411}^4:=\mathbb{P}\left[(|4+5\rangle_{A} ;|1\rangle_{a}]\right.,\\A_{412}^4=\mathbb{P}\left[|4-5\rangle_{A} ;|1\rangle_{a}]\right.\}
$$
\end{multline*}
to distinguish the states $|\phi\prime_{31,32}\rangle$. Also, depending on $B_{42}^3$  Alice performs,
\begin{multline*}
$$
\mathcal{A}^4_{42}\equiv\{A_{421}^4:=\mathbb{P}\left[|4\rangle_{A};|1\rangle_{a}]\right., A_{422}^4=\mathbb{P}\left[|5\rangle_{A} ;|1\rangle_{a}]\right.\}.
$$
\end{multline*}
The outcomes $A_{421}^4$ and $A_{422}^4$ will isolate $|\phi\prime_{33,34}\rangle$ and $|\phi\prime_{35,36}\rangle$ respectively.
Next, depending on the outcome $B_{51}^3$ of the previous round, Alice performs the measurement,
\begin{multline*}
$$
\mathcal{A}^4_{51}\equiv\{A_{511}^4:=\mathbb{P}\left[(|1+2\rangle_{A} ;|0\rangle_{a}]\right.,\\A_{512}^4=\mathbb{P}\left[|1-2\rangle_{A} ;|0\rangle_{a}]\right.,\\A_{513}^4:=\mathbb{P}\left[(|3+4\rangle_{A} ;|0\rangle_{a}]\right.,\\A_{514}^4=\mathbb{P}\left[|3-4\rangle_{A} ;|0\rangle_{a}]\right.,\\A_{515}^4:=\mathbb{P}\left[(|5\rangle_{A} ;|0\rangle_{a}]\right.\}
$$
\end{multline*}
Now the outcomes $A_{511}^4, B_{512}^4, B_{513}^4, B_{514}^4, B_{515}^4$ distinguish $|\phi\prime_{7,8,9,10,11}\rangle$ respectively. Again, depending on $B_{53}^3$ Alice performs,
\begin{multline*}
$$
\mathcal{A}^4_{54}\equiv\{A_{541}^4:=\mathbb{P}\left[|1\rangle_{A} ;|0\rangle_{a}\right],A_{542}^4:=\mathbb{I}-A_{541}^4\}.
$$
\end{multline*}
If the outcome $A_{541}^4$ clicks, it isolates $|\phi\prime_{12,13}\rangle$ and if the outcome $A_{542}^4$ clicks, it isolates $|\phi\prime_{17,18,19,20,21,22,25,26,27}\rangle$.\\
{\bfseries Round 5.} Bob performs the measurement,\\
depending on the outcome $A_{421}^4$ of the previous round, Bob performs the measurement,
\begin{multline*}
$$
\mathcal{B}^5_{421}\equiv\{B_{4211}^5:=\mathbb{P}\left[(|4+5\rangle_{B} ;|1\rangle_{b}]\right.,\\B_{4212}^5=\mathbb{P}\left[|4-5\rangle_{B} ;|1\rangle_{b}]\right.\}
$$
\end{multline*}
The outcomes $B_{4211}^5$ and $B_{4212}^5$ successfully  distinguish the states $|\phi\prime_{33,34}\rangle$. Next depending on $A_{422}^4$ Bob performs the measurement,
\begin{multline*}
$$
\mathcal{B}^5_{422}\equiv\{B_{4221}^5:=\mathbb{P}\left[|4\rangle_{B} ;|1\rangle_{b}]\right.,B_{4222}^5=\mathbb{P}\left[|5\rangle_{B} ;|1\rangle_{b}]\right.\}
$$
\end{multline*}
The outcomes $B_{4221}^5$ and $B_{4222}^5$ successfully  distinguish the states $|\phi\prime_{35,36}\rangle$. Next, depending on the outcome $A_{541}^4$ of the previous round, Bob performs the measurement,
\begin{multline*}
$$
\mathcal{B}^5_{541}\equiv\{B_{5411}^5:=\mathbb{P}\left[(|1+2\rangle_{B} ;|0\rangle_{b}]\right.,\\B_{5412}^5=\mathbb{P}\left[|1-2\rangle_{B} ;|0\rangle_{b}]\right.\}
$$
\end{multline*}
The outcomes $B_{5411}^5$ and $B_{5412}^5$ successfully distinguish the states $|\phi\prime_{12,13}\rangle$. Next depending on $A_{542}^4$ Bob performs,
\begin{multline*}
$$
\mathcal{B}^5_{542}\equiv\{B_{5421}^5:=\mathbb{P}\left[|1\rangle_{B} ;|0\rangle_{b}]\right.,\\B_{5422}^5=\mathbb{P}\left[|2\rangle_{B} ;|0\rangle_{b}]\right.+\left[|3\rangle_{B} ;|1\rangle_{b}]\right.\}
$$
\end{multline*}
If the outcome $B_{5421}^5$ clicks it isolates $|\phi\prime_{17,18,19,20}\rangle$ and if the outcome $B_{5422}^5$ clicks it isolates $|\phi\prime_{21,22,25,26,27}\rangle$.\\
{\bfseries Round 6.} Alice performs the measurement,\\
Next, depending on the outcome $B_{5421}^5$ of the previous round, Alice performs the measurement,
\begin{multline*}
$$
\mathcal{A}^6_{5421}\equiv\{A_{54211}^6:=\mathbb{P}\left[|2+3\rangle_{A};|0\rangle_{a}]\right.,\\A_{54212}^6:=\mathbb{P}\left[|2-3\rangle_{A};|0\rangle_{a}]\right.,\\A_{54213}^6:=\mathbb{P}\left[|4+5\rangle_{A};|0\rangle_{a}]\right.,\\A_{54214}^6:=\mathbb{P}\left[|4-5\rangle_{A};|0\rangle_{a}]\right.\}
$$
\end{multline*}
The outcomes $A_{54211}^6$,$A_{54212}^6$,$A_{54213}^6$ and $A_{54214}^6$ successfully  distinguish the states $|\phi\prime_{17,18,19,20}\rangle$. Next, depending on $B_{5422}^5$ Alice performs the measurement,
\begin{multline*}
$$
\mathcal{A}_{5422}^6\equiv\{A_{54221}^6=\mathbb{P}\left[|2\rangle_{A} ;|0+1\rangle_{a}]\right.,\\A_{54222}^6=\mathbb{P}\left[|2\rangle_{A} ;|0-1\rangle_{a}]\right.,\\A_{54223}^6:=\mathbb{P}\left[|3+4\rangle_{A};|0\rangle_{a}]\right.,\\A_{54224}^6:=\mathbb{P}\left[|3-4\rangle_{A};|0\rangle_{a}]\right.,\\A_{54225}^6:=\mathbb{P}\left[|5\rangle_{A};|0\rangle_{a}]\right.\}
$$
\end{multline*}
The outcomes $A_{54221}^6$,$A_{54222}^6$,$A_{54223}^6$,$A_{54224}^6$ and $A_{54225}^6$ successfully  distinguish the states $|\phi\prime_{21,22,25,26,27}\rangle$. Hence the proof is complete.\(\blacksquare \)\\\\
{\large\textbf{Proof of Proposition 4.}}\\
First of all, let us assume that the state with one-ebit of entanglement shared between Alice, Bob be $|\psi\rangle_{ab}$. Therefore the initial state shared among them is $\left|\phi\right\rangle_{AB}\otimes\left|\psi\right\rangle_{ab}$,
where $\left|\phi\right\rangle$ is one of the state from set of equations (2).

Round $1 .$ Bob performs a measurement
\begin{multline*}
$$
\mathcal{B}\equiv\{B_1:=\mathbb{P}\left[(|0\rangle,|1\rangle,|2\rangle,\ldots|\frac{d}{2}-1\rangle)_{B} ;|0\rangle_{b}\right]+ \\ \mathbb{P}[(|\frac{d}{2}\rangle,|\frac{d}{2}+1\rangle,\ldots|d-1\rangle)_{B} ;|1\rangle_{b}],\\
B_2:=\mathbb{I}-B_1\}
$$
\end{multline*}
Suppose the outcomes corresponding to $B_1$ click. The resulting postmeasurement states are therefore
\begin{multline*}
$$
\left|\phi_{i+1, i+2}\right\rangle \rightarrow|0\rangle_{A}|i\pm(i+1)\rangle_{B}|00\rangle_{ab}, i=0,2, \ldots,\frac{d}{2}-3,\;\;\;\;\;\;\;\;\;\\
\left|\phi_{\frac{d}{2},\frac{d}{2}+1}\right\rangle \rightarrow|0\rangle_{A}|{\frac{d}{2}}-1\rangle_{B}|00\rangle_{ab}\pm|0\rangle_{A}|{\frac{d}{2}}\rangle_{B}|11\rangle_{ab},\;\;\;\;\;\;\;\;\;\;\;\;\;\;\;\;\;\;\;\;\;\;\;\;\;\;\;\;\;\;\;\;\;\;\;\;\;\\
\left|\phi_{i+1, i+2}\right\rangle \rightarrow|0\rangle_{A}|i\pm(i+1)\rangle_{B}|11\rangle_{ab}, i=\frac{d}{2}+1,\frac{d}{2}+3, \ldots,d-2,\;\;\;\;\;\;\;\;\;\\
\left|\phi_{d+i, d+i+1}\right\rangle\rightarrow|i\pm(i+1)\rangle_{A}|0\rangle_{B}|00\rangle_{ab}, i=1,3, \ldots, d-3,\;\;\;\;\;\\
\left|\phi_{2d-1}\right\rangle \rightarrow|d-1\rangle_{A}|0\rangle_{B}|00\rangle_{ab},\;\;\;\;\;\;\;\;\;\;\;\;\;\;\;\;\;\;\;\;\;\;\;\;\;\;\;\;\;\;\;\;\;\;\;\;\;\;\;\;\;\;\;\;\;\;\;\;\;\;\\
\left|\phi_{2d+i-1,2d+i}\right\rangle \rightarrow|1\rangle_{A}|i\pm(i+1)\rangle_{B}|00\rangle_{ab}, i=1,3, \ldots, \frac{d}{2}-2,\;\;\;\\
\left|\phi_{2d+i-1,2d+i}\right\rangle \rightarrow|1\rangle_{A}|i\pm(i+1)\rangle_{B}|11\rangle_{ab}, i=\frac{d}{2}, \frac{d}{2}+2, \ldots,d-3,\;\;\;\\	
\left|\phi_{3d-2}\right\rangle \rightarrow|1\rangle_{A}|d-1\rangle_{B}|11\rangle_{ab},\;\;\;\;\;\;\;\;\;\;\;\;\;\;\;\;\;\;\;\;\;\;\;\;\;\;\;\;\;\;\;\;\;\;\;\;\;\;\;\;\;\;\;\;\;\;\;\;\;\;\\
\left|\phi_{3d+i-3,3d+i-2}\right\rangle \rightarrow|i\pm (i+1)\rangle_{A}|1\rangle_{B}|00\rangle_{ab}, i=2,4, \ldots, d-2,\\
\left|\phi_{4d+i-5,4d+i-4}\right\rangle =|2\rangle_{A}|i\pm (i+1)\rangle_{B}|00\rangle_{ab}, i=2,4, \ldots,\frac{d}{2}-3,\\
|\phi_{4d+\frac{d}{2}-6,4d+\frac{d}{2}-5}\rangle \rightarrow|2\rangle_{A}|\frac{d}{2}-1\rangle_{B}|00\rangle_{ab}\pm|2\rangle_{A}|\frac{d}{2}\rangle_{B}|11\rangle_{ab},\;\;\;\;\;\;\\
\left|\phi_{4d+i-5,4d+i-4}\right\rangle\rightarrow|2\rangle_{A}|i\pm (i+1)\rangle_{B}|00\rangle_{ab}, i=\frac{d}{2}+1,\ldots,d-2,\\
\left|\phi_{5d+i-8,5d+i-7}\right\rangle\rightarrow|i\pm (i+1)\rangle_{A}|2\rangle_{B}|00\rangle_{ab}, i=3,5, \ldots, d-3,\\
\left|\phi_{6d-9}\right\rangle\rightarrow|d-1\rangle_{A}|2\rangle_{B}|00\rangle_{ab},\;\;\;\;\;\;\;\;\;\;\;\;\;\;\;\;\;\;\;\;\;\;\;\;\;\;\;\;\;\;\;\;\;\;\;\;\;\;\;\;\;\;\;\;\;\;\;\;\;\;\;\\
\left|\phi_{6d+i-11,6d+i-10}\right\rangle\rightarrow|3\rangle_{A}|i \pm(i+1)\rangle_{B}|00\rangle_{ab}, i=3,5,\ldots,\frac{d}{2}-2,\\
\left|\phi_{6d+i-11,6d+i-10}\right\rangle\rightarrow|3\rangle_{A}|i \pm(i+1)\rangle_{B}|00\rangle_{ab}, i=\frac{d}{2},\ldots,d-3,,\\
\left|\phi_{7d-12}\right\rangle\rightarrow|3\rangle_{A}|d-1\rangle_{B}|11\rangle_{ab},\;\;\;\;\;\;\;\;\;\;\;\;\;\;\;\;\;\;\;\;\;\;\;\;\;\;\;\;\;\;\;\;\;\;\;\;\;\;\;\;\;\;\;\;\;\;\;\;\;\;\\
\vdots\;\;\;\;\;\;\;\;\;\;\;\;\;\;\;\;\;\;\;\;\;\;\;\;\;\;\;\;\;\;\;\;\;\;\;\;\;\;\;\;\;\;\;\;\;\;\;\;\;\;\\
\left|\phi_{d^2-5,d^2-4}\right\rangle\rightarrow|(d-2)\pm(d-1)\rangle_{A}|d-3\rangle_{B}|11\rangle_{ab},\;\;\;\;\;\;\;\;\;\;\;\;\;\;\;\;\;\;\;\;\;\;\;\;\;\;\;\;\;\;\;\;\;\;\;\;\\
\left|\phi_{d^2-3,d^2-2}\right\rangle\rightarrow|d-2\rangle_{A}|(d-2)\pm(d-1)\rangle_{B}|11\rangle_{ab},\;\;\;\;\;\;\;\;\;\;\;\;\;\;\;\;\;\;\;\;\;\;\;\;\;\;\;\;\;\;\;\;\;\;\;\;\\
\left|\phi_{d^2-1}\right\rangle\rightarrow|d-1\rangle_{A}|d-2\rangle_{B}|11\rangle_{ab},\;\;\;\;\;\;\;\;\;\;\;\;\;\;\;\;\;\;\;\;\;\;\;\;\;\;\;\;\;\;\;\;\;\;\;\;\;\;\;\;\;\;\;\;\;\;\;\;\;\;\\
\left|\phi_{d^2}\right\rangle\rightarrow|d-1\rangle_{A}|d-1\rangle_{B}|11\rangle_{ab},\;\;\;\;\;\;\;\;\;\;\;\;\;\;\;\;\;\;\;\;\;\;\;\;\;\;\;\;\;\;\;\;\;\;\;\;\;\;\;\;\;\;\;\;\;\;\;\;\;\;\\
$$
\end{multline*}
Next Alice and Bob will do a sequence of measurements to distinguish those states as we have done in
previous one. \(\blacksquare \)\\\\\\
{\large\textbf{Proof of Proposition 5.}}\\
Here also we first assume that the state with 2-ebits of entanglement shared between Alice, Bob be $|\psi_1\rangle_{a_1b_1} \otimes|\psi_2\rangle_{a_2b_2}$, where each of $|\psi_1\rangle_{a_1b_1}$ and $|\psi_2\rangle_{a_2b_2}$ have one-ebit of entanglement. Therefore the initial state shared among them is $$\left|\phi\right\rangle_{AB}\otimes\left|\psi_1\right\rangle_{a_1b_1}\otimes\left|\psi_2\right\rangle_{a_2b_2}$$
where $\left|\phi\right\rangle$ is one of the state from set of equations (1).\\
{\bfseries Round 1.} Bob performs a measurement,
\begin{multline*}
$$
\mathcal{B}\equiv\left\{B_1:=\mathbb{P}\left[(|0\rangle,|1\rangle,|2\rangle)_{B} ;|0\rangle_{b}\right]+\mathbb{P}\left[(|3\rangle,|4\rangle,|5\rangle)_{B} ;|1\rangle_{b}\right]\right.,\\\;\;\;\;\;\;\;\;\;\;\;\;\;\;\;\;\;\;\;\;\;\;\;\;\;\;\;\;\;\;\;\;\;\;\;\;\;\;\;\;\;\;\;\;\;\;\;\;\;\;\;\;\;\;\;\;\;\;\;\;\;\;\;\;
B_2:=\mathbb{I}-B_1\}
$$\\
$$
\mathcal{C}\equiv\left\{C_1:=\mathbb{P}\left[(|0\rangle,|4\rangle,|5\rangle)_{B} ;|0\rangle_{b}\right]+\mathbb{P}\left[(|1\rangle,|2\rangle,|3\rangle)_{B} ;|1\rangle_{b}\right]\right.,\;\;\;\;\;\;\;\\\;\;\;\;\;\;\;\;\;\;\;\;\;\;
C_2:=\mathbb{I}-C_1\}
$$
\end{multline*}
Suppose the outcomes corresponding to $B_1$ and $C_1$ clicks. The resulting postmeasurement states are therefore
\begin{multline*}
$$
\left|\phi\prime_{1,2}\right\rangle \rightarrow\left|0\rangle_{A}|0\rangle_{B}|00\rangle_{a_{1}b_{1}}|00\rangle_{a_{2}b_{2}}\right.\pm\;\;\;\;\;\;\;\;\;\;\;\;\;\;\;\;\;\;\;\;\;\;\;\;\;\;\;\;\;\;\;\;\;\\\;\;\;\;\;\;\;\;\;\;\;\;\;\;\;\;\;\;\;\;\;\;\;\left|0\rangle_{A}|1\rangle_{B}|00\rangle_{a_{1}b_{1}}|11\rangle_{a_{2}b_{2}}\right.,\\
\left|\phi\prime_{3,4}\right\rangle \rightarrow\left|0\rangle_{A}|2\rangle_{B}|00\rangle_{a_{1}b_{1}}|11\rangle_{a_{2}b_{2}}\right.\pm\;\;\;\;\;\;\;\;\;\;\;\;\;\;\;\;\;\;\;\;\;\;\;\;\;\;\;\;\;\;\;\;\;\\\;\;\;\;\;\;\;\;\;\;\;\;\;\;\;\;\;\;\;\;\;\;\;\;\left|0\rangle_{A}|3\rangle_{B}|11\rangle_{a_{1}b_{1}}|11\rangle_{a_{2}b_{2}}\right.,\\
\left|\phi\prime_{5,6}\right\rangle \rightarrow\left|0\rangle_{A}|4\pm5\rangle_{B}|11\rangle_{a_{1}b_{1}}|00\rangle_{a_{2}b_{2}}\right.,\;\;\;\;\;\;\;\;\;\;\;\;\;\;\;\;\;\;\;\;\;\;\;\;\;\;\;\;\;\;\;\\
\left|\phi\prime_{7,8}\right\rangle \rightarrow\left|1\pm2\rangle_{A}|0\rangle_{B}|00\rangle_{a_{1}b_{1}}|00\rangle_{a_{2}b_{2}}\right.,\;\;\;\;\;\;\;\;\;\;\;\;\;\;\;\;\;\;\;\;\;\;\;\;\;\;\;\;\;\;\\
\left|\phi\prime_{9,10}\right\rangle \rightarrow\left|3\pm4\rangle_{A}|0\rangle_{B}|00\rangle_{a_{1}b_{1}}|00\rangle_{a_{2}b_{2}}\right.,\;\;\;\;\;\;\;\;\;\;\;\;\;\;\;\;\;\;\;\;\;\;\;\;\;\;\;\;\;\\
\left|\phi\prime_{11}\right\rangle \rightarrow\left|5\rangle_{A}|0\rangle_{B}|00\rangle_{a_{1}b_{1}}|00\rangle_{a_{2}b_{2}}\right.,\;\;\;\;\;\;\;\;\;\;\;\;\;\;\;\;\;\;\;\;\;\;\;\;\;\;\;\;\;\;\;\;\;\;\;\;\;\\
\left|\phi\prime_{12,13}\right\rangle \rightarrow\left|1\rangle_{A}|1\pm2\rangle_{B}|00\rangle_{a_{1}b_{1}}|11\rangle_{a_{2}b_{2}}\right.,\;\;\;\;\;\;\;\;\;\;\;\;\;\;\;\;\;\;\;\;\;\;\;\;\;\;\;\;\;\;\\
\left|\phi\prime_{14,15}\right\rangle \rightarrow\left|1\rangle_{A}|3\rangle_{B}|11\rangle_{a_{1}b_{1}}|11\rangle_{a_{2}b_{2}}\right.\pm\;\;\;\;\;\;\;\;\;\;\;\;\;\;\;\;\;\;\;\;\;\;\;\;\;\;\;\;\;\;\;\;\;\\\;\;\;\;\;\;\;\;\;\;\;\;\;\;\;\;\;\;\;\;\;\;\;\;\;\left|1\rangle_{A}|4\rangle_{B}|11\rangle_{a_{1}b_{1}}|00\rangle_{a_{2}b_{2}}\right.,\\
\left|\phi\prime_{16}\right\rangle \rightarrow\left|1\rangle_{A}|5\rangle_{B}|00\rangle_{a_{1}b_{1}}|11\rangle_{a_{2}b_{2}}\right.,\;\;\;\;\;\;\;\;\;\;\;\;\;\;\;\;\;\;\;\;\;\;\;\;\;\;\;\;\;\;\;\;\;\;\;\;\;\\
\left|\phi\prime_{17,18}\right\rangle \rightarrow\left|2\pm3\rangle_{A}|1\rangle_{B}|00\rangle_{a_{1}b_{1}}|11\rangle_{a_{2}b_{2}}\right.,\;\;\;\;\;\;\;\;\;\;\;\;\;\;\;\;\;\;\;\;\;\;\;\;\;\;\;\;\;\;\\
\left|\phi\prime_{19,20}\right\rangle \rightarrow\left|4\pm5\rangle_{A}|1\rangle_{B}|00\rangle_{a_{1}b_{1}}|11\rangle_{a_{2}b_{2}}\right.,\;\;\;\;\;\;\;\;\;\;\;\;\;\;\;\;\;\;\;\;\;\;\;\;\;\;\;\;\;\;\\
\left|\phi\prime_{21,22}\right\rangle \rightarrow\left|2\rangle_{A}|2\rangle_{B}|00\rangle_{a_{1}b_{1}}|11\rangle_{a_{2}b_{2}}\right.\pm\;\;\;\;\;\;\;\;\;\;\;\;\;\;\;\;\;\;\;\;\;\;\;\;\;\;\;\;\;\;\;\;\;\\\;\;\;\;\;\;\;\;\;\;\;\;\;\;\;\;\;\;\;\;\;\;\;\;\;\;\;\left|2\rangle_{A}|3\rangle_{B}|11\rangle_{a_{1}b_{1}}|11\rangle_{a_{2}b_{2}}\right.,\\
\left|\phi\prime_{23,24}\right\rangle \rightarrow\left|2\rangle_{A}|4\pm5\rangle_{B}|11\rangle_{a_{1}b_{1}}|00\rangle_{a_{2}b_{2}}\right.,\;\;\;\;\;\;\;\;\;\;\;\;\;\;\;\;\;\;\;\;\;\;\;\;\;\;\;\;\;\\
\left|\phi\prime_{25,26}\right\rangle \rightarrow\left|3\pm4\rangle_{A}|2\rangle_{B}|00\rangle_{a_{1}b_{1}}|11\rangle_{a_{2}b_{2}}\right.,\;\;\;\;\;\;\;\;\;\;\;\;\;\;\;\;\;\;\;\;\;\;\;\;\;\;\;\;\;\\
\left|\phi\prime_{27}\right\rangle \rightarrow\left|5\rangle_{A}|2\rangle_{B}|00\rangle_{a_{1}b_{1}}|11\rangle_{a_{2}b_{2}}\right.,\;\;\;\;\;\;\;\;\;\;\;\;\;\;\;\;\;\;\;\;\;\;\;\;\;\;\;\;\;\;\;\;\;\;\;\;\;\\
\left|\phi\prime_{28,29}\right\rangle \rightarrow\left|3\rangle_{A}|3\rangle_{B}|11\rangle_{a_{1}b_{1}}|11\rangle_{a_{2}b_{2}}\right.\pm\;\;\;\;\;\;\;\;\;\;\;\;\;\;\;\;\;\;\;\;\;\;\;\;\;\;\;\;\;\;\;\;\;\;\;\;\;\;\;\;\;\\\;\;\;\;\;\;\;\;\;\;\;\;\;\;\;\;\;\;\;\;\;\;\;\;\;\;\;\;\;\left|3\rangle_{A}|4\rangle_{B}|11\rangle_{a_{1}b_{1}}|00\rangle_{a_{2}b_{2}}\right.,\\
\left|\phi\prime_{30}\right\rangle \rightarrow\left|3\rangle_{A}|5\rangle_{B}|11\rangle_{a_{1}b_{1}}|00\rangle_{a_{2}b_{2}}\right.,\;\;\;\;\;\;\;\;\;\;\;\;\;\;\;\;\;\;\;\;\;\;\;\;\;\;\;\;\;\;\;\;\;\;\;\;\;\\
\left|\phi\prime_{31,32}\right\rangle \rightarrow\left|4\pm5\rangle_{A}|3\rangle_{B}|11\rangle_{a_{1}b_{1}}|11\rangle_{a_{2}b_{2}}\right.,\;\;\;\;\;\;\;\;\;\;\;\;\;\;\;\;\;\;\;\;\;\;\;\;\;\;\;\;\;\\
\left|\phi\prime_{33,34}\right\rangle \rightarrow\left|4\rangle_{A}|4\pm5\rangle_{B}|11\rangle_{a_{1}b_{1}}|00\rangle_{a_{2}b_{2}}\right.,\;\;\;\;\;\;\;\;\;\;\;\;\;\;\;\;\;\;\;\;\;\;\;\;\;\;\;\;\;\\
\left|\phi\prime_{35}\right\rangle \rightarrow\left|5\rangle_{A}|4\rangle_{B}|11\rangle_{a_{1}b_{1}}|00\rangle_{a_{2}b_{2}}\right.,\;\;\;\;\;\;\;\;\;\;\;\;\;\;\;\;\;\;\;\;\;\;\;\;\;\;\;\;\;\;\;\;\;\;\;\;\;\\
\left|\phi\prime_{36}\right\rangle \rightarrow\left|5\rangle_{A}|5\rangle_{B}|11\rangle_{a_{1}b_{1}}|00\rangle_{a_{2}b_{2}}\right.,\;\;\;\;\;\;\;\;\;\;\;\;\;\;\;\;\;\;\;\;\;\;\;\;\;\;\;\;\;\;\;\;\;\;\;\;\;\\
$$
\end{multline*}
{\bfseries Round 2.} Alice performs a measurement,
\begin{multline*}
$$
\mathcal{A}^2\equiv\{A_1^2:=\mathbb{P}\left[|0\rangle_{A} ;(|0\rangle,|1\rangle)_{a_1};(|0\rangle,|1\rangle)_{a_2}]\right.,\\A_2^2:=\mathbb{P}\left[|1+2\rangle_{A} ;|0\rangle_{a_1};|0\rangle_{a_2}]\right.,\\A_3^2:=\mathbb{P}\left[|1-2\rangle_{A} ;|0\rangle_{a_1};|0\rangle_{a_2}]\right.,\\A_4^2:=\mathbb{P}\left[|3+4\rangle_{A} ;|0\rangle_{a_1};|0\rangle_{a_2}]\right.,\\A_5^2:=\mathbb{P}\left[|3-4\rangle_{A} ;|0\rangle_{a_1};|0\rangle_{a_2}]\right.,\\A_6^2:=\mathbb{P}\left[|5\rangle_{A} ;|0\rangle_{a_1};|0\rangle_{a_2}]\right.,\\A_7^2:=\mathbb{P}\left[|1\rangle_{A} ;|1\rangle_{a_1};(|0\rangle,|1\rangle)_{a_2}]\right.+\\\mathbb{P}\left[|1\rangle_{A} ;|0\rangle_{a_1};|1\rangle_{a_2}]\right.,\\A_8^2:=\mathbb{P}\left[|2\rangle_{A} ;|1\rangle_{a_1};|0\rangle_{a_2}]\right.,\\A_9^2:=\mathbb{P}\left[|3\rangle_{A} ;|1\rangle_{a_1};(|0\rangle,|1\rangle)_{a_2}]\right.\\A_{10}^2:=\mathbb{P}\left[|4+5\rangle_{A} ;|1\rangle_{a_1};|1\rangle_{a_2}]\right.,\\A_{11}^2:=\mathbb{P}\left[|4-5\rangle_{A} ;|1\rangle_{a_1};|1\rangle_{a_2}]\right.,\\A_{12}^2:=\mathbb{P}\left[|4\rangle_{A} ;|1\rangle_{a_1};|0\rangle_{a_2}]\right.,\\A_{13}^2:=\mathbb{P}\left[|5\rangle_{A} ;|1\rangle_{a_1};|0\rangle_{a_2}]\right.,A_{14}^2:=\mathbb{I}-\sum A_i^2\}
$$
\end{multline*}
If the outcomes corresponding to $A_1^2$ click. The resulting post measurement states are therefore $	\left|\phi\prime_{1,2,3,4,5,6}\right\rangle $. The outcomes $A_2^2,A_3^2,A_4^2,A_5^2,A_6^2$ respectively identified the states $\left|\phi\prime_{7,8,9,10,11}\right\rangle$. The outcomes corresponding to $A_7^2$ isolates the states $\left|\phi\prime_{12,13,14,15,16}\right\rangle$. The outcome $A_8^2$ eliminates $ \left|\phi\prime_{23,24}\right\rangle $. If the outcomes corresponding to $A_9^2$ click. The resulting post measurement states are therefore $	\left|\phi\prime_{28,29,30}\right\rangle $. The outcomes $A_{10}^2,A_{11}^2$ respectively identified the states $\left|\phi\prime_{31,32}\right\rangle$. The outcomes corresponding to $A_{12}^2$ isolates the states $\left|\phi\prime_{33,34}\right\rangle$. The outcomes corresponding to $A_{13}^2$ isolates the states $\left|\phi\prime_{35,36}\right\rangle$. If the outcomes corresponding to $A_{14}^2$ click, it isolates the remaining states $	\left|\phi\prime_{17,18,19,20,21,22,25,26,27}\right\rangle $.\\
{\bfseries Round 3.} Bob performs the measurement,\\
depending on the outcome $A_1^2$ of the previous round, Bob performs four-outcomes projective measurement,
\begin{multline*}
$$
\mathcal{B}^3_{1}\equiv\{B_{11}^3:=\mathbb{P}\left[|0\rangle_{B} ;|0\rangle_{b_1};|0\rangle_{b_2}]\right.\\+\mathbb{P}\left[|1\rangle_{B} ;|0\rangle_{b_1};|1\rangle_{b_2}]\right.,\\B_{12}^3:=\mathbb{P}\left[|2\rangle_{B} ;|0\rangle_{b_1};|1\rangle_{b_2}]\right.\\+\mathbb{P}\left[|3\rangle_{B} ;|1\rangle_{b_1};|1\rangle_{b_2}]\right.,\\B_{13}^3=\mathbb{P}\left[|4+5\rangle_{B} ;|1\rangle_{b_1};|0\rangle_{b_2}]\right.,\\B_{14}^3:=\mathbb{P}\left[|4-5\rangle_{B} ;|1\rangle_{b_1};|0\rangle_{b_2}]\right.\}
$$
\end{multline*}
The outcomes  $B_{11}^3$ and $B_{12}^3$ isolates  $|\phi\prime_{1,2}\rangle$ and $|\phi\prime_{3,4}\rangle$ respectively. Whereas the other two outcomes $B_{13}^3$ and $B_{14}^3$ successfully identifies $|\phi\prime_{5}\rangle$ and $|\phi\prime_{6}\rangle$ respectively. Depending on the $A_7^2$, Bob performs four-outcomes projective measurement,
\begin{multline*}
$$
\mathcal{B}^3_{7}\equiv\{B_{71}^3=\mathbb{P}\left[|1+2\rangle_{B} ;|0\rangle_{b_1};|1\rangle_{b_2}]\right.,\\B_{72}^3:=\mathbb{P}\left[|1-2\rangle_{B} ;|0\rangle_{b_1};|1\rangle_{b_2}]\right.,\\B_{73}^3:=\mathbb{P}\left[|3\rangle_{B} ;|1\rangle_{b_1};|1\rangle_{b_2}]\right.\\+\mathbb{P}\left[|4\rangle_{B} ;|1\rangle_{b_1};|0\rangle_{b_2}]\right.,\\B_{74}^3:=\mathbb{P}\left[|5\rangle_{B} ;|1\rangle_{b_1};|0\rangle)_{b_2}]\right.\}
$$
\end{multline*}
The outcomes $B_{71}^3$, $B_{72}^3$ and $B_{74}^3$ identifies $|\phi\prime_{12}\rangle$, $|\phi\prime_{13}\rangle$ and $|\phi\prime_{16}\rangle$ respectively. Whereas the outcome  $B_{73}^3$ isolates the remaining two states $|\phi\prime_{14,15}\rangle$. The post measurement reduced states corresponding to the outcome $A_8^2$ in the previous round was $ \left|\phi\prime_{23,24}\right\rangle $, which can be perfectly distinguished by Bob by projecting onto $|4\pm5\rangle_B$.\\
Depending on the outcome $A_9^2$ of the previous round, Bob performs two-outcomes projective measurement,
\begin{multline*}
$$
\mathcal{B}^3_{9}\equiv\{B_{91}^3:=\mathbb{P}\left[|3\rangle_{B} ;|1\rangle_{b_1};|1\rangle_{b_2}]\right.\\+\mathbb{P}\left[|4\rangle_{B} ;|1\rangle_{b_1};|0\rangle_{b_2}]\right.,\\B_{92}^3:=\mathbb{P}\left[|5\rangle_{B} ;|1\rangle_{b_1};|0\rangle)_{b_2}]\right.\}
$$
\end{multline*}
The outcome $B_{92}^3$ will identify the state $|\phi\prime_{30}\rangle$. Whereas the outcome  $B_{91}^3$ isolates the remaining two states $|\phi\prime_{28,29}\rangle$. Each of the outcomes $A_{12}^2$ and $A_{13}^2$ in the previous round isolates two states $( \left|\phi\prime_{33,34}\right\rangle, \left|\phi\prime_{35,36}\right\rangle )$, which can be perfectly distinguished by Bob by projecting onto $|4\pm5\rangle_B$ and $\{|4\rangle,|5\rangle\}_B$.\\
Now if the outcome $A_{14}^2$ occur in the previous round, Bob will make two-outcomes projective measurement,
\begin{multline*}
$$
\mathcal{B}^3_{14}\equiv\{B_{141}^3:=\mathbb{P}\left[|2\rangle_{B} ;|0\rangle_{b_1};|1\rangle_{b_2}]\right.\\+\mathbb{P}\left[|3\rangle_{B} ;|1\rangle_{b_1};|1\rangle_{b_2}]\right.,\\B_{142}^3:=\mathbb{P}\left[|1\rangle_{B} ;|0\rangle_{b_1};|1\rangle)_{b_2}]\right.\}
$$
\end{multline*}
The outcome corresponding to $B_{141}^3$ will isolate the states $( \left|\phi\prime_{17,18,19,20}\right\rangle$ and the outcome corresponding to $B_{142}^3$ will isolate the states $( \left|\phi\prime_{21,22,25,26,27}\right\rangle$.\\
{\bfseries Round 4.} Alice performs the measurement\\
As each of the outcomes $B_{11}^3$, $B_{12}^3$, $B_{73}^3$ and $B_{91}^3$ in the previous round isolates two states. Therefore it can be distinguished by Walgate et.al.(Phys. Rev. Lett. 85, 4972 (2000)) results in either case.\\
The post measurement reduced states corresponding to the outcome $B_{141}^3$ in the previous round was $ \left|\phi\prime_{17,18,19,20}\right\rangle $, which can be perfectly distinguished by Bob by projecting onto $|2\pm3\rangle_B$ and $|4\pm5\rangle_B$ respectively.\\
Depending on the outcome $B_{142}^3$ of the previous round, Alice performs five-outcomes projective measurement,
\begin{multline*}
$$
\mathcal{A}^4_{142}\equiv\{A_{1421}^4=\mathbb{P}\left[|3+4\rangle_{A} ;|0\rangle_{a_1};|1\rangle_{a_2}]\right.,\\A_{1422}^4:=\mathbb{P}\left[|3-4\rangle_{A} ;|0\rangle_{a_1};|1\rangle_{a_2}]\right.,\\A_{1423}^4:=\mathbb{P}\left[|2\rangle_{A} ;|0+1\rangle_{a_1};|1\rangle_{a_2}]\right.,\\A_{1424}^4:=\mathbb{P}\left[|2\rangle_{A} ;|0-1\rangle_{a_1};|1\rangle_{a_2}]\right.,\\A_{1425}^3:=\mathbb{P}\left[|5\rangle_{A} ;|0\rangle_{a_1};|1\rangle_{a_2}]\right.\}
$$
\end{multline*}
Hence the outcomes $A_{1421}^4$, $A_{1422}^4$, $A_{1423}^4$, $A_{1424}^4$ and $A_{1425}^4$ respectively identifies the states $|\phi\prime_{21,22,25,26,27}\rangle$. This completes the proof.\(\blacksquare \)\\\\
{\large\textbf{Proof of Proposition 6.}}\\
First of all let us assume that the state with 2-ebits of entanglement shared between Alice, Bob be $|\psi_1\rangle_{a_1b_1} \otimes|\psi_2\rangle_{a_2b_2}$, where each of $|\psi_1\rangle_{a_1b_1}$ and $|\psi_2\rangle_{a_2b_2}$ have one-ebit of entanglement. Therefore the initial state shared among them is $\left|\phi\right\rangle_{AB}\otimes\left|\psi_1\right\rangle_{a_1b_1}\otimes\left|\psi_2\right\rangle_{a_2b_2}$,
where $\left|\phi\right\rangle$ is one of the state from set of equations (2).\\
Round $1 .$ Bob performs a measurement
\begin{multline*}
$$
\mathcal{B}\equiv\{B_1:=\mathbb{P}[(|0\rangle,|1\rangle,\ldots,|\frac{d}{2}-1\rangle)_{B} ;|0\rangle_{b}]+\\\mathbb{P}[(|\frac{d}{2}\rangle,|\frac{d}{2}+1\rangle,\ldots,|d-1\rangle)_{B} ;|1\rangle_{b}],\\\;\;\;\;\;\;\;\;\;\;\;\;\;\;\;\;\;\;\;\;\;\;\;\;\;\;\;\;\;\;\;\;\;\;\;\;\;\;\;\;\;\;\;\;\;\;\;\;\;\;\;\;\;\;\;\;\;\;\;\;\;\;\;\;
B_2:=\mathbb{I}-B_1\}
$$\\
$$
\mathcal{C}\equiv\{C_1:=\mathbb{P}[(|0\rangle,|\frac{d}{2}+1\rangle,\ldots,|d-1\rangle)_{B} ;|0\rangle_{b}]+\\\mathbb{P}[(|1\rangle,|2\rangle,\ldots,|\frac{d}{2}\rangle)_{B} ;|1\rangle_{b}],\;\;\;\;\;\;\;\\\;\;\;\;\;\;\;\;\;\;\;\;\;\;
C_2:=\mathbb{I}-C_1\}
$$
\end{multline*}
Suppose the outcomes corresponding to $B_1$ and $C_1$ click. The resulting postmeasurement states are therefore
\begin{multline*}
$$
\left|\phi\prime_{1,2}\right\rangle \rightarrow\left|0\rangle_{A}|0\rangle_{B}|00\rangle_{a_{1}b_{1}}|00\rangle_{a_{2}b_{2}}\right.\pm\;\;\;\;\;\;\;\;\;\;\;\;\;\;\;\;\;\;\;\;\;\;\;\;\;\;\;\;\;\;\;\;\;\\\;\;\;\;\;\;\;\;\;\;\;\;\;\;\;\;\;\;\;\;\;\;\;\left|0\rangle_{A}|1\rangle_{B}|00\rangle_{a_{1}b_{1}}|11\rangle_{a_{2}b_{2}}\right.,\\
\left|\phi\prime_{i+1,i+2}\right\rangle \rightarrow\left|0\rangle_{A}|i\rangle_{B}|00\rangle_{a_{1}b_{1}}|11\rangle_{a_{2}b_{2}}\right.\pm\;\;\;\;\;\;\;\;\;\;\;\;\;\;\;\;\;\;\;\;\;\;\;\;\;\;\;\;\;\;\;\;\;\\\;\;\;\;\;\;\;\;\;\;\;\;\;\;\;\;\left|0\rangle_{A}|i+1\rangle_{B}|00\rangle_{a_{1}b_{1}}|11\rangle_{a_{2}b_{2}}\right., i=2,4, \ldots,\frac{d}{2}-2,\\
\left|\phi\prime_{\frac{d}{2}+1,\frac{d}{2}+2}\right\rangle \rightarrow|0\rangle_{A}|\frac{d}{2}\rangle_{B}|11\rangle_{a_{1}b_{1}}|11\rangle_{a_{2}b_{2}}\pm\;\;\;\;\;\;\;\;\;\;\;\;\;\;\;\;\;\;\;\;\;\;\;\;\;\;\;\;\;\;\;\;\;\\\;\;\;\;\;\;\;\;\;\;\;\;\;\;\;\;|0\rangle_{A}|\frac{d}{2}+1\rangle_{B}|11\rangle_{a_{1}b_{1}}|00\rangle_{a_{2}b_{2}},\\
\left|\phi\prime_{\frac{d}{2}+,6}\right\rangle \rightarrow\left|0\rangle_{A}|4\pm5\rangle_{B}|11\rangle_{a_{1}b_{1}}|00\rangle_{a_{2}b_{2}}\right.,\;\;\;\;\;\;\;\;\;\;\;\;\;\;\;\;\;\;\;\;\;\;\;\;\;\;\;\;\;\;\;\\
\left|\phi\prime_{7,8}\right\rangle \rightarrow\left|1\pm2\rangle_{A}|0\rangle_{B}|00\rangle_{a_{1}b_{1}}|00\rangle_{a_{2}b_{2}}\right.,\;\;\;\;\;\;\;\;\;\;\;\;\;\;\;\;\;\;\;\;\;\;\;\;\;\;\;\;\;\;\\
\left|\phi\prime_{9,10}\right\rangle \rightarrow\left|3\pm4\rangle_{A}|0\rangle_{B}|00\rangle_{a_{1}b_{1}}|00\rangle_{a_{2}b_{2}}\right.,\;\;\;\;\;\;\;\;\;\;\;\;\;\;\;\;\;\;\;\;\;\;\;\;\;\;\;\;\;\\
\left|\phi\prime_{11}\right\rangle \rightarrow\left|5\rangle_{A}|0\rangle_{B}|00\rangle_{a_{1}b_{1}}|00\rangle_{a_{2}b_{2}}\right.,\;\;\;\;\;\;\;\;\;\;\;\;\;\;\;\;\;\;\;\;\;\;\;\;\;\;\;\;\;\;\;\;\;\;\;\;\;\\
\left|\phi\prime_{12,13}\right\rangle \rightarrow\left|1\rangle_{A}|1\pm2\rangle_{B}|00\rangle_{a_{1}b_{1}}|11\rangle_{a_{2}b_{2}}\right.,\;\;\;\;\;\;\;\;\;\;\;\;\;\;\;\;\;\;\;\;\;\;\;\;\;\;\;\;\;\;\\
\left|\phi\prime_{14,15}\right\rangle \rightarrow\left|1\rangle_{A}|3\rangle_{B}|11\rangle_{a_{1}b_{1}}|11\rangle_{a_{2}b_{2}}\right.\pm\;\;\;\;\;\;\;\;\;\;\;\;\;\;\;\;\;\;\;\;\;\;\;\;\;\;\;\;\;\;\;\;\;\\\;\;\;\;\;\;\;\;\;\;\;\;\;\;\;\;\;\;\;\;\;\;\;\;\;\left|1\rangle_{A}|4\rangle_{B}|11\rangle_{a_{1}b_{1}}|00\rangle_{a_{2}b_{2}}\right.,\\
\left|\phi\prime_{16}\right\rangle \rightarrow\left|1\rangle_{A}|5\rangle_{B}|00\rangle_{a_{1}b_{1}}|11\rangle_{a_{2}b_{2}}\right.,\;\;\;\;\;\;\;\;\;\;\;\;\;\;\;\;\;\;\;\;\;\;\;\;\;\;\;\;\;\;\;\;\;\;\;\;\;\\
\left|\phi\prime_{17,18}\right\rangle \rightarrow\left|2\pm3\rangle_{A}|1\rangle_{B}|00\rangle_{a_{1}b_{1}}|11\rangle_{a_{2}b_{2}}\right.,\;\;\;\;\;\;\;\;\;\;\;\;\;\;\;\;\;\;\;\;\;\;\;\;\;\;\;\;\;\;\\
\left|\phi\prime_{19,20}\right\rangle \rightarrow\left|4\pm5\rangle_{A}|1\rangle_{B}|00\rangle_{a_{1}b_{1}}|11\rangle_{a_{2}b_{2}}\right.,\;\;\;\;\;\;\;\;\;\;\;\;\;\;\;\;\;\;\;\;\;\;\;\;\;\;\;\;\;\;\\
\left|\phi\prime_{21,22}\right\rangle \rightarrow\left|2\rangle_{A}|2\rangle_{B}|00\rangle_{a_{1}b_{1}}|11\rangle_{a_{2}b_{2}}\right.\pm\;\;\;\;\;\;\;\;\;\;\;\;\;\;\;\;\;\;\;\;\;\;\;\;\;\;\;\;\;\;\;\;\;\\\;\;\;\;\;\;\;\;\;\;\;\;\;\;\;\;\;\;\;\;\;\;\;\;\;\;\;\left|2\rangle_{A}|3\rangle_{B}|11\rangle_{a_{1}b_{1}}|11\rangle_{a_{2}b_{2}}\right.,\\
\left|\phi\prime_{23,24}\right\rangle \rightarrow\left|2\rangle_{A}|4\pm5\rangle_{B}|11\rangle_{a_{1}b_{1}}|00\rangle_{a_{2}b_{2}}\right.,\;\;\;\;\;\;\;\;\;\;\;\;\;\;\;\;\;\;\;\;\;\;\;\;\;\;\;\;\;\\
\left|\phi\prime_{25,26}\right\rangle \rightarrow\left|3\pm4\rangle_{A}|2\rangle_{B}|00\rangle_{a_{1}b_{1}}|11\rangle_{a_{2}b_{2}}\right.,\;\;\;\;\;\;\;\;\;\;\;\;\;\;\;\;\;\;\;\;\;\;\;\;\;\;\;\;\;\\
\left|\phi\prime_{27}\right\rangle \rightarrow\left|5\rangle_{A}|2\rangle_{B}|00\rangle_{a_{1}b_{1}}|11\rangle_{a_{2}b_{2}}\right.,\;\;\;\;\;\;\;\;\;\;\;\;\;\;\;\;\;\;\;\;\;\;\;\;\;\;\;\;\;\;\;\;\;\;\;\;\;\\
\left|\phi\prime_{28,29}\right\rangle \rightarrow\left|3\rangle_{A}|3\rangle_{B}|11\rangle_{a_{1}b_{1}}|11\rangle_{a_{2}b_{2}}\right.\pm\;\;\;\;\;\;\;\;\;\;\;\;\;\;\;\;\;\;\;\;\;\;\;\;\;\;\;\;\;\;\;\;\;\;\;\;\;\;\;\;\;\\\;\;\;\;\;\;\;\;\;\;\;\;\;\;\;\;\;\;\;\;\;\;\;\;\;\;\;\;\;\left|3\rangle_{A}|4\rangle_{B}|11\rangle_{a_{1}b_{1}}|00\rangle_{a_{2}b_{2}}\right.,\\
\left|\phi\prime_{30}\right\rangle \rightarrow\left|3\rangle_{A}|5\rangle_{B}|11\rangle_{a_{1}b_{1}}|00\rangle_{a_{2}b_{2}}\right.,\;\;\;\;\;\;\;\;\;\;\;\;\;\;\;\;\;\;\;\;\;\;\;\;\;\;\;\;\;\;\;\;\;\;\;\;\;\\
\left|\phi\prime_{31,32}\right\rangle \rightarrow\left|4\pm5\rangle_{A}|3\rangle_{B}|11\rangle_{a_{1}b_{1}}|11\rangle_{a_{2}b_{2}}\right.,\;\;\;\;\;\;\;\;\;\;\;\;\;\;\;\;\;\;\;\;\;\;\;\;\;\;\;\;\;\\
\left|\phi\prime_{33,34}\right\rangle \rightarrow\left|4\rangle_{A}|4\pm5\rangle_{B}|11\rangle_{a_{1}b_{1}}|00\rangle_{a_{2}b_{2}}\right.,\;\;\;\;\;\;\;\;\;\;\;\;\;\;\;\;\;\;\;\;\;\;\;\;\;\;\;\;\;\\
\left|\phi\prime_{35}\right\rangle \rightarrow\left|5\rangle_{A}|4\rangle_{B}|11\rangle_{a_{1}b_{1}}|00\rangle_{a_{2}b_{2}}\right.,\;\;\;\;\;\;\;\;\;\;\;\;\;\;\;\;\;\;\;\;\;\;\;\;\;\;\;\;\;\;\;\;\;\;\;\;\;\\
\left|\phi\prime_{36}\right\rangle \rightarrow\left|5\rangle_{A}|5\rangle_{B}|11\rangle_{a_{1}b_{1}}|00\rangle_{a_{2}b_{2}}\right.,\;\;\;\;\;\;\;\;\;\;\;\;\;\;\;\;\;\;\;\;\;\;\;\;\;\;\;\;\;\;\;\;\;\;\;\;\;\\
$$
\end{multline*}
Next Alice and Bob will do a sequence of measurements to distinguish those states as we have done in
previous one. \(\blacksquare \)\\\\

\end{document}